\documentstyle[revtex,eqsecnum]{aps}
\begin{document}

\begin{title}
Meson mass spectrum from relativistic equations in configuration
space
\end{title}

\author{Peter C. Tiemeijer and J.A. Tjon}

\begin{instit}
Institute for Theoretical Physics, University of Utrecht,
3508 TA Utrecht, The Netherlands.
\end{instit}

\begin{abstract}
A method is described for solving relativistic quasi-potential
equations in configuration space. The
Blankenbecler-Sugar-Logunov-Tavkhelidze and an equal-time equation,
both relativistic covariant two-body equations containing the full
Dirac structure of positive and negative energy states, are studied
in detail. These equations are solved for a system of two
constituent quarks interacting through a potential consisting of a
one-gluon exchange part with running coupling constant plus a linear
confining potential which is mostly scalar and partly vector, and
the spectrum of all light and heavy mesons is calculated.
\end{abstract}


\section{INTRODUCTION}

It is rather surprising how well the non-relativistic constituent
quark model does describe the masses of the known mesons and baryons
\cite{isgur}. However, non-relativistic models with or without
relativistic
corrections can be criticized easily of having an inaccurate
treatment of the quark dynamics when high momentum processes or
light quarks are considered. It is therefore interesting to study
relativistic covariant extensions of the constituent quark model,
which are still likely to reproduce the hadron spectroscopy with
similar agreement as the older models, but have a more reliable
dynamical framework at extreme momenta. In a previous
paper~\cite{peter} we studied the pi and rho meson in such an
extension, which was based on a covariant equal-time approximation
to the Bethe-Salpeter equation. It was shown that the known high
momentum behavior of the electromagnetic form factors could be
reproduced within this model.

In this present work we investigate two different relativistic
covariant approximations to the Bethe-Salpeter equation, namely the
equal-time approximation~\cite{wallace} and the
Blankenbecler-Sugar-Logunov-Tavkhelidze approximation
\cite{bslt,cooper} and calculate the meson mass spectrum. The
ingredients of the calculations are constituent quarks
interacting through a instantaneous flavor-independent
phenomenological quark-antiquark potential. The potential consists
of a Coulomb-like one-gluon-exchange part with running coupling
constant and of a linear plus constant confining part which is a
mixture of scalar and vector character. For the vector part of the
potential two different gauges, Feynman and Coulomb, are examined.

The Bethe-Salpeter equation (BSE) is the relativistic covariant
generalization of the Lippmann-Schwinger equation for a two-quark
bound state and can be written as
\begin{equation}
S^{-1}(P,p)\psi(P,p)=-\int \!\frac{d^4p'}{i(2\pi)^4} \,V(P,p-p')\,
\psi(P,p'),
\end{equation}
where $P$ and $p$ are the total and relative momenta. Two new
features are introduced as compared to the non-relativistic
equation. The first one is that the bound state wave function
$\psi(P,p)$ and the potential $V(P,p-p')$ now also depend on the
fourth component of the relative momentum. The second is that for
each quark the number of degrees of freedom is doubled due to the
introduction of the negative energy states. These extensions make
the numerical work necessary for solving the BSE highly non trivial.
In order to reduce the complexity one may remove all or some of the
negative energy states, or one may eliminate the dependence on the
fourth component of the relative momentum. The Lorentz invariance
and the structure of the negative energy states can be kept if this
component is eliminated.

There have been various suggestions how to approximate the relative
energy dependence of the BSE~\cite{brownjack}, most of which are
based on a prescription for modifying the two body propagator in
such a way that the relative energy is forced to some fixed value.
This modification of the propagator is equivalent to a modification
of the potential and the resulting wave equations are commonly
called quasi-potential equations. Here we study two of them. The
first one is the Blankenbecler-Sugar-Logunov-Tavkhelidze (BSLT)
approximation, originally proposed by Logunov and Tavkhelidze, and
Blankenbecler and Sugar~\cite{bslt}, and extended to unequal masses
by Cooper and Jennings~\cite{cooper}. The second is a two-body Dirac
equation due to Wallace and Mandelzweig~\cite{wallace}, which may be
viewed as an extension of the equal-time approximation of
Salpeter~\cite{salpeter}. These quasi-potential equations are
normally formulated in momentum space because many expressions in
the equations depend on functions like $\sqrt{p_i^2+m_i^2}$.
However, the confining potential in momentum space is highly
singular at zero relative momentum $q$. From $V(q\!=\!0)=\int\!
dx^3\, V(x)$ and the monotonical increasement of the confining
potential $V(x)$ one easily finds that $V(q)$ diverges stronger than
$q^{-3}$ for small momenta. Special care must be given when solving
such equations~\cite{murota,vary,maung,gross}. This work uses a
different approach. Using well-chosen representations of the
quasi-potential equations we found that they can be Fourier
transformed easily to configuration space, thus encompassing all
problems associated with the confining potential.

This paper is organized as follows. In the next section we  discuss
the derivation of the BSLT and ET equations and how they can be
Fourier transformed from momentum space to configuration space. The
partial wave projection of the equations is given in appendix A. In
section III we describe the interaction and its implications for the
short distance behavior of the corresponding solutions. The presence
of the Coulomb-like potential makes the wave function singular at
zero relative distance. This sets an upper bound for fixed coupling
constants. The short distance behavior of the $J^P\!=\!0^-$ states
(e.g. the pion) is analyzed in detail in appendix B. In  section IV
we describe the method of solution  of the resulting
integro-differential equations. Also we study the parameter
dependence and global behavior of relativistic effects in the meson
mass spectrum.  Section V discusses the Regge-trajectories which
follow from the mass spectra of the BSLT and ET equations. The
slopes of these trajectories are found to be dependent on the degree
of admixture of vector character in the confining potential. If it
becomes too large the potential becomes repulsive at long distances
in some channels making the solutions of the wave equation unbound.
This gives an upper bound for the degree of the vector character in
the confining potential.  In section VI the full meson spectrum is
discussed as obtained  from the various quasi-potential equations
and in the final section some concluding remarks are made.

\section{THE QUASI-POTENTIAL EQUATIONS}
\noindent
For the full BSE it is of no importance how the total and relative
momenta are defined. However, the prescriptions for the
quasi-potential approximation depend on the definition of the
relative momentum. Let us follow Wightman and
Garding~\cite{wightman} and define the total and relative momenta
$P$ and $p$ for a two particle system by
\begin{equation}
\begin{array}{rcl} P & = & p_1 + p_2, \rule[-4mm]{0ex}{4mm} \\
              p & = & \beta(s)p_1 - \alpha(s)p_2, \end{array}
\ \ \ \ \ \ \
\begin{array}{rcl} p_1 & = & \alpha(s) P +p,\rule[-4mm]{0ex}{4mm} \\
              p_2 & = & \beta(s) P -p, \end{array}
\end{equation}
and
\begin{equation}
\alpha(s)=\frac{s+m_1^2-m_2^2}{2s}, \ \ \ \ \ \ \
\beta(s)=\frac{s-m_1^2+m_2^2}{2s},
\end{equation}
where $s=-P^2=P_0^2-\mbox{\boldmath$P$}^2$, and $p_1$ and $p_2$
denote the momenta of the two particles. In the center of mass
system, where $P=(M,\mbox{\boldmath$0$})$, they read
\begin{eqnarray}
p_1 & = & (\frac{M^2+m_1^2-m_2^2}{2M}+p_0,\mbox{\boldmath$p$}) =
(E_1
+p_0,\mbox{\boldmath$p$}), \\ p_2 & = &
(\frac{M^2-m_1^2+m_2^2}{2M}-p_0,-\mbox{\boldmath$p$})
 =
(E_2 -p_0,-\mbox{\boldmath$p$}).
\end{eqnarray}
This definition of $p$ has the advantage that in the limit of one
infinitely heavy particle, say $m_2 \rightarrow \infty$, that $E_1
\rightarrow M - m_2$. Now $M- m_2$ is precisely the total energy
associated with the light particle, so this implies that the
relative energy $p_0$ goes to zero. It turns out that this last
property is essential for the quasi-potential approximations to have
the correct one-body limit, i.e., that they reduce the one-body
Dirac equation in the limit of one infinitely heavy particle.

\subsection{Blankenbecler-Sugar-Logunov-Tavkhelidze equation}
\noindent
The propagator of the Bethe-Salpeter equation is
\begin{eqnarray}
S(p_1,p_2)&=&(\not\! p_1-m_1)^{-1}(\not\! p_2-m_2)^{-1}
\nonumber \rule[-4mm]{0ex}{4mm} \\
& = & (\not\! p_1+m_1)(\not\! p_2+m_2)\  S^{scalar}(p_1,p_2).
\end{eqnarray}
The pole-structure of this propagator is contained in the scalar
part of the propagator $S^{scalar}(p_1,p_2)$. In the approximation
first proposed by Logunov and Tavkhelidze, and Blankenbecler and
Sugar \cite{bslt}, it is assumed that the pole-structure of the
scalar part may be approximated by means of a dispersion relation
according to
\begin{equation}
S^{scalar} \!\rightarrow S_{BSLT}^{scalar}=2\pi i
\int_{(m_1+m_2)^2}^{\infty} \!\!\!\!ds'\frac{f(s',s)}{s'-s}
\delta^+\! \left( [\alpha(s') P'+p]^2+m_1^2 \right)
\delta^+\! \left( [\beta(s') P'-p]^2+m_2^2 \right),
\end{equation}
where $P'=\sqrt{s'/s} P$, and $f(s',s)$ is any function that
satisfies $f(s,s)=1$. Performing the integral gives in the cm system
\begin{equation}
S_{BSLT}^{scalar} =
\frac{f((\omega_1+\omega_2)^2,s)}{(\omega_1+\omega_2)^2-s} \
\frac{\omega_1+\omega_2}{2\omega_1\omega_2} \ \delta(p_0),
\end{equation}
with $\omega_i=\sqrt{\mbox{\boldmath $p$}^2+m_i^2}$. Taking for
$f(s',s)$ the simple form proposed by Cooper and Jennings
\cite{cooper}, which yields the proper one-body limit in the case
that one of the masses goes to infinity, one finds
\begin{equation}
S_{BSLT}^{scalar} =
-\frac{1}{s-(\omega_1+\omega_2)^2} \
 \frac{1}{s-(\omega_1-\omega_2)^2} \
\frac{2s}{\omega_1+\omega_2} \ \delta(p_0),
\end{equation}
and
\begin{equation}
S_{BSLT}(P,p) = (\not\!{\tilde p}_1+m_1)(\not\!{\tilde p}_2+m_2)\
S_{BSLT}^{scalar}(P,p).
\end{equation}
The $\tilde p_1$ and $\tilde p_2$ follow from $p_1$ and $p_2$ by
putting the relative energy to zero, thus $\tilde p_1 =
(E_1,\mbox{\boldmath$p$})$ and $\tilde p_2 =
(E_2,-\mbox{\boldmath$p$})$. Note that this propagator may be cast
in the simple forms
\begin{eqnarray}
S_{BSLT} & = &
-\frac{4s}{\left[s-(\omega_1+\omega_2)^2\right]
\left[s-(\omega_1-\omega_2)^2\right]}
\ \frac{\delta(p_0)}{2(\omega_1+\omega_2)}
\ (\not\!{\tilde p}_1+m_1)(\not\!{\tilde p}_2+m_2) \hspace{1.5cm} \\
 & = & \rule[-6.5mm]{0ex}{15mm}
 \frac{\delta(p_0)}{2(\omega_1+\omega_2)}
\ \frac{\not\!{\tilde p}_1+m_1}{\not\!{\tilde p}_2-m_2}
\label{qq1}\\
 & = &
 \frac{\delta(p_0)}{2(\omega_1+\omega_2)}
\ \frac{\not\!{\tilde p}_2+m_2}{\not\!{\tilde p}_1-m_1} \label{qq2}
\end{eqnarray}
It is instructive to project the BSLT propagator upon positive and
negative energy states. Using
\begin{equation}
\not\! p_i +m_i =
(\omega_i + p_{0i}-i\varepsilon)\;\Lambda_i^+(\mbox{\boldmath$p$}_i)
- (\omega_i -
p_{0i}-i\varepsilon)\;\Lambda_i^-(\mbox{\boldmath$p$}_i),
\label{qq13}
\end{equation}
with
\begin{equation}
\Lambda_i^{\rho_i}(\mbox{\boldmath$p$}_i) =
\frac{ \rho_i (-\mbox{\boldmath$\gamma$}^{(i)} \cdot
\mbox{\boldmath$p$}_i + m_i)
+\omega_i \gamma_0^{(i)} }{2 \omega_i},
\end{equation}
and introducing $\Lambda^{\rho_1
\rho_2}=\Lambda_1^{\rho_1}(\mbox{\boldmath$p$})
\Lambda_2^{\rho_2}(-\mbox{\boldmath$p$})$ gives
\begin{eqnarray}
S_{BSLT} =
\frac{\delta(p_0)}{2(\omega_1+\omega_2)}\ \frac{1}{G}
&&\left[(\omega_1+E_1)(\omega_2+E_2)\Lambda^{++}
-(\omega_1+E_1)(\omega_2-E_2)\Lambda^{+-} \right.
\nonumber \\
&&
\left.
-(\omega_1-E_1)(\omega_2+E_2)\Lambda^{-+}
+(\omega_1-E_1)(\omega_2-E_2)\Lambda^{--}\right],
\hspace{2em}\label{qq4}
\end{eqnarray}
where $G=\omega_1^2-E_1^2=\omega_2^2-E_2^2$. This form clearly shows
that the BSLT propagator is time-reversal invariant. It satisfies
the condition that it does not change form if one interchanges
simultaneously the energy labels $\rho_i \rightarrow -\rho_i$ and
the particle energies $E_i \rightarrow -E_i$. If the potential is
also time-reversal invariant then the BSLT equation has the property
that if a solution exists with mass $M$ then there will also be a
solution at $-M$, which is identical to the solution at $M$ but with
positive energy states changed to negative energy states and vice
versa. These solutions at $-M$ correspond of course to the
antiparticles of the solutions at $M$.

Since we want to solve the BSLT equation $S_{BSLT}^{-1} \psi= - V
\psi$ in configuration space we
have to Fourier transform it. In order to bring this equation in a
form which may be easily transformed we multiply it by
\begin{equation}
D=\frac{\not\!{\tilde p}_1 + \not\!{\tilde p}_2 + m_1 +
m_2}{2(\omega_1+\omega_2)}
\label{qq25}
\end{equation}
which, using eq.~(\ref{qq1}) and~(\ref{qq2}), gives
\begin{equation}
(\not\!{\tilde p}_1 + \not\!{\tilde p}_2 - m_1 - m_2)
\psi(P,\mbox{\boldmath$p$}) =
\frac{1}{2(\omega_1+\omega_2)}
(\not\!{\tilde p}_1 + \not\!{\tilde p}_2 + m_1 + m_2)
\int \frac{d\mbox{\boldmath$p$}'}{(2\pi)^3} \
V(\mbox{\boldmath$p$}'-\mbox{\boldmath$p$})
\psi(P,\mbox{\boldmath$p$}').
\label{qq26}
\end{equation}
In this form it is most clearly visible that in the limit that one
of the particles masses goes to infinity, say $m_2 \rightarrow
\infty$, the BSLT equation reduces to the Dirac equation
\begin{equation}
(\not\!{\tilde p}_1 - m_1) \psi(\mbox{\boldmath$p$}_1) =
\int \frac{d\mbox{\boldmath$p$}_1'}{(2\pi)^3} \
V(\mbox{\boldmath$p$}_1'-\mbox{\boldmath$p$}_1)
\psi(\mbox{\boldmath$p$}'_1).
\end{equation}
The Fourier transform of the BSLT equation now reads
\begin{eqnarray}
\lefteqn{\left[ i\mbox{\boldmath$\gamma$}^{(1)} \cdot
\mbox{\boldmath$\nabla$} + E_1 \gamma_0^{(1)}
-i\mbox{\boldmath$\gamma$}^{(2)} \cdot \mbox{\boldmath$\nabla$} +
E_2
\gamma_0^{(2)} - m_1 - m_2 \right]
\psi(P, \mbox{\boldmath$x$}) = }
\nonumber \\
& & \rule{0ex}{5ex}
\int\! d\mbox{\boldmath$x$}'\  \frac{Z_{BSLT}(\mbox{\boldmath$x$}' -
\mbox{\boldmath$x$})}{2(m_1+m_2)}
\nonumber \\
& & \rule{0ex}{5ex}
\times
\left[i\mbox{\boldmath$\gamma$}^{(1)} \cdot \mbox{\boldmath$\nabla$}
+
E_1 \gamma_0^{(1)}
-i\mbox{\boldmath$\gamma$}^{(2)} \cdot \mbox{\boldmath$\nabla$} +
E_2
\gamma_0^{(2)} + m_1 + m_2 \right] V(\mbox{\boldmath$x$}')
\psi(P,\mbox{\boldmath$x$}').
\hspace{2em}
\label{qq6}
\end{eqnarray}
Due to the relativistic phase space factor in the BSLT propagator, a
non-locality occurs in the relativistic equation, which is contained
in the function
\begin{equation}
Z_{BSLT}(\mbox{\boldmath$R$}) =
\int \!\frac{d\mbox{\boldmath$p$}}{(2\pi)^3}\
\frac{m_1+m_2}{\omega_1+\omega_2}
e^{\textstyle i\mbox{\boldmath$p$}\cdot\mbox{\boldmath$R$}}.
\end{equation}
This integral can be found by rewriting
\begin{equation}
\frac{m_1+m_2}{\omega_1+\omega_2} =
\frac{1}{m_1-m_2}
\left[ (m_1^2-\mbox{\boldmath$\nabla$}_R^2) \frac{1}{\omega_1}
-(m_2^2-\mbox{\boldmath$\nabla$}_R^2) \frac{1}{\omega_2}\right],
\end{equation}
and using the Fourier transform of $m_i/\omega_i$
\begin{equation}
Z_i(\mbox{\boldmath$R$}) =
\int \!\frac{d\mbox{\boldmath$p$}}{(2\pi)^3}\  e^{\textstyle
i\mbox{\boldmath$p$}\cdot\mbox{\boldmath$R$}}\
\frac{m_i}{\omega_i} = \frac{m^2_i}{2\pi^2 R} \;K_1(m_i R).
\label{qq15}
\end{equation}
$K_1$ is the modified Bessel function of the second kind of order
one. The non-locality becomes
\begin{eqnarray}
\lefteqn{Z_{BSLT}(R) =
\frac{1}{2\pi^2 R^2} \ \frac{1}{m_1-m_2}}
\nonumber \\
 &&\times \rule{0ex}{5ex}
\left[-m_1^2\left\{ K_0(m_1 R) +\frac{2}{m_1 R} K_1(m_1 R) \right\}
+m_2^2\left\{ K_0(m_2 R) +\frac{2}{m_2 R} K_1(m_2 R) \right\}
\right].
\hspace{2em}
\label{qq5}
\end{eqnarray}
which is invariant under the interchange $m_1 \leftrightarrow m_2$.
Both $Z_{BSLT}(R)$ and $Z_i(R)$ behave like $1/R^2$ for short
distances. If the two quarks have equal masses then the
non-localities become identical, $Z_{BSLT}(R)=Z_1(R)=Z_2(R)$. In
order to be solved, the BSLT equation is projected on eight basis
states with definite parity and total angular momentum $J$. These
basis states, the partial wave analysis and the resulting eight
coupled integro-differential equations are discussed in
appendix~\ref{appa1}. As example, the reader may look at
Eq.~(\ref{qq23}) to see the explicit BSLT equation for the pion.

\subsection{Equal-time approximation}

In the equal-time (ET) approximation the assumption is made that the
q\=q-potential $V$ does not depend on the relative energy in the cm
system. This simplifies the BSE considerably. One may then introduce
the equal-time wave function $\psi(\mbox{\boldmath$p$}) = 1/(2\pi i)
\int\! dp_0\, \psi(p_0,\mbox{\boldmath$p$})$, or
$\psi(\mbox{\boldmath$x$}) = \psi(x_0\!=\!0,\mbox{\boldmath$x$})$,
and integrate the BSE on both sides over $p_0$ to obtain an equation
for the ET wave function:
\begin{equation}
\psi(\mbox{\boldmath$p$}) = -\left[\int \frac {dp_0}{2\pi i}
S(p_1,p_2) \right] \int \frac{d\mbox{\boldmath$p$}}{(2\pi)^3}
V(\mbox{\boldmath$p$} - \mbox{\boldmath$p$}')
\psi(\mbox{\boldmath$p$}').
\label{qq12}
\end{equation}
For the sake of clarity the derivations in this section are
formulated in the cm frame and therefore the ET approach does not
show a manifestly relativistic covariant form. Yet it can be
formulated in fully manifestly relativistic covariant way, see
Ref.~\cite{wallace}.

The ET equation as its stands in Eq.~(\ref{qq12}) does not reduce to
the Dirac equation if one of the particles masses becomes infinitely
heavy and so it does not have the correct one-body limit. However,
as has been pointed out by Wallace and Mandelzweig~\cite{wallace},
the ET approximation can be improved considerably by adding a
propagator $S_{cross}$ to the integral between square brackets,
which represents the propagation of the two particles within the
crossed box diagram. With this extra term the ET propagator exhibits
the one-body limit, and furthermore the contributions of the crossed
diagrams are approximately taken into account. The propagator is now
taken as
\begin{equation}
S_{ET} = \int \frac{dp_0}{2\pi i}
\left[ S(p_1,p_2) + S_{cross}(p_1,p_2) \right],
\end{equation}
where in the cm system
\begin{equation}
S_{cross}(p_1,p_2)=S(p_1,p_2^{cross}),\ \ \
p_2^{cross}=(E_2+p_0,-\mbox{\boldmath$p$}).
\end{equation}
The only difference between $p_2$ and $p_2^{cross}$ is the sign in
front of the relative energy. The origin of the different signs for
$p_0$ can easily be understood from the different directions of flow
of relative energy in the uncrossed and crossed diagrams. A detailed
justification for this crossed box contribution based on the eikonal
approximation has been given in ref.~\cite{wallace}. The integral
over $p_0$ can be evaluated easily by using Eq.~(\ref{qq13}) and
expressing the propagators in positive and negative energy
projections. This gives
\begin{eqnarray}
S_{ET}&=&\frac{\Lambda^{++}}{\omega_1+\omega_2-E_1-E_2}
      -\frac{\Lambda^{-+}}{\omega_1+\omega_2+E_1-E_2}
\nonumber \rule[-6.5mm]{0ex}{6.5mm} \\
  & & -\frac{\Lambda^{+-}}{\omega_1+\omega_2-E_1+E_2}
      +\frac{\Lambda^{--}}{\omega_1+\omega_2+E_1+E_2},
\label{qq14}
\end{eqnarray}
The terms containing $\Lambda^{++}$ and $\Lambda^{--}$ are
well-known and stem from the $p_0$-integration over $S$. They were
first used by Salpeter~\cite{salpeter}. The other two terms result
from the $p_0$-integration over $S_{cross}$. This form shows that
the ET propagator is time-reversal invariant, similarly as the BSLT
propagator. The most important difference between the BSLT and ET
approximation is the degree to which the propagation of negative
energy states is suppressed as compared to the propagation of the
positive energy states. For particles of equal mass
($\omega_1=\omega_2=\omega$, and $E_1=E_2=E$), the ratio of the
propagation for positive and negative states is for the ET case
\begin{equation}
\frac{S_{ET}^{++}}{S_{ET}^{--}} =
\frac{\omega + E}{\omega - E},
\label{qq28}
\end{equation}
whereas Eq.~(\ref{qq4}) gives for the BSLT case
\begin{equation}
\frac{S_{BSLT}^{++}}{S_{BSLT}^{--}} =
\left(\frac{\omega + E}{\omega - E}\right)^2.
\label{qq29}
\end{equation}
Therefore we can expect a larger admixture of negative states in the
ET solutions than in the BSLT solutions. Note that this ratio is
always positive for the BSLT propagator whereas it can be negative
for the ET propagator. For the full BSE it is always positive at
$p_0=0$.

Writing out the projection operators $\Lambda^{\rho_1\rho_2}$ gives
\begin{eqnarray}
S_{ET}^{-1} & = &
\frac{1}{\omega_1} \left[-\mbox{\boldmath$\gamma$}^{(1)} \cdot
\mbox{\boldmath$p$} + m_1\right]
\left[\mbox{\boldmath$\gamma$}^{(2)} \cdot \mbox{\boldmath$p$} -
\gamma_0^{(2)} E_2+ m_2 \right]
\nonumber \rule[-6.5mm]{0ex}{6.5mm} \\
 & & +\frac{1}{\omega_2} \left[
 -\mbox{\boldmath$\gamma$}^{(1)} \cdot \mbox{\boldmath$p$} -
\gamma_0^{(1)} E_1+ m_1 \right]
\left[ \mbox{\boldmath$\gamma$}^{(2)} \cdot \mbox{\boldmath$p$} +
m_2\right]
\end{eqnarray}
The ET bound state wave equation $S_{ET}^{-1} \psi = -V \psi$ may
now be transformed easily to configuration space and becomes
\begin{equation}
\int \!d\mbox{\boldmath$x$}' \left[
\frac{Z_1(\mbox{\boldmath$x$}-\mbox{\boldmath$x$}')}{m_1} \;{{\cal
S}}_1(\mbox{\boldmath$x$}')
+\frac{Z_2(\mbox{\boldmath$x$}-\mbox{\boldmath$x$}')}{m_2} \;{{\cal
S}}_2(\mbox{\boldmath$x$}')
\right] \psi (\mbox{\boldmath$x$}') =-V(\mbox{\boldmath$x$}) \psi
(\mbox{\boldmath$x$}),
\end{equation}
with
\begin{eqnarray}
{{\cal S}}_1(\mbox{\boldmath$x$}) & = &
\left[i\mbox{\boldmath$\gamma$}^{(1)} \cdot
\mbox{\boldmath$\nabla$}+m_1\right]
\left[-i\mbox{\boldmath$\gamma$}^{(2)} \cdot
\mbox{\boldmath$\nabla$}
-\gamma_0^{(2)} E_2+m_2 \right],
\label{qq16}
\\
\rule{0ex}{3ex} {{\cal S}}_2(\mbox{\boldmath$x$}) & = &
\left[i\mbox{\boldmath$\gamma$}^{(1)} \cdot \mbox{\boldmath$\nabla$}
-\gamma_0^{(1)} E_1 +m_1\right]
\left[-i\mbox{\boldmath$\gamma$}^{(2)} \cdot
\mbox{\boldmath$\nabla$}+m_2\right],
\label{qq17}
\end{eqnarray}
and the non-localities $Z_i$ have already been given in
Eq.~(\ref{qq15}). The ET equation can be solved after projecting it
on basis states with definite parity and total angular momentum $J$.
The resulting eight coupled integro-differential equations can be
found in appendix~\ref{appa2}.

\section{INTERACTION}

It is commonly accepted that the interaction between the two quarks
consists of a short-range part describing the one-gluon-exchange
(OGE) potential and a infinitely rising long-range part responsible
for the confinement of the quarks~\cite{lucha}. We use
\begin{equation}
V(x)=-\frac{\alpha(x)}{x} \gamma_\mu^{(1)} \gamma^{\mu (2)}
+(\kappa x + c) \left[ (1-\varepsilon) 1^{(1)} 1^{(2)}
+ \varepsilon \gamma_\mu^{(1)} \gamma^{\mu (2)}\right].
\label{qq7}
\end{equation}
(The color factor $4/3$ of the expectation value of the OGE in the
meson color wave function has been absorbed in the definition of
$\alpha$.) The OGE potential is a pure vector interaction. The
confining potential is commonly believed to be purely scalar.
However, we choose a confining potential which is mainly scalar, but
it can have a fraction $\varepsilon$ of vector confinement.
Asymptotic freedom requires that for short distances the OGE
coupling constant decreases logarithmically as $\alpha(x) \sim
\alpha_0 /\ln(x_0/x)$ where $\alpha_0=8 \pi/(33-2n_F)$ and $x_0 =
e^{-\gamma}/\Lambda_{QCD}$
\cite{lucha}. Richardson~\cite{richardson} has given an elegant
prescription for a running coupling constant which for small
distances reproduces the correct asymptotic freedom and which gives
for large distances a linearly rising potential. We do not use this
prescription since it can not specify the vector and scalar
character of the potential. At large distances the confining
interaction dominates and the exact behavior of $\alpha(x)$ becomes
of little importance. If one assumes that the coupling constant
grows to some saturation value $\alpha_{sat}$ then a smooth
interpolation between the short and long range is given
by~\cite{levine}
\begin{equation}
\alpha_I(x)=\alpha_0 \
\ln^{-1}\left[
\frac{x_0}{x} \exp(-\mu x) + \exp\left(\frac{\alpha_0}{\alpha_{sat}}
\right)\right].
\label{qq27}
\end{equation}
Here the typical range of the running coupling regime is controlled
by the parameter $\mu$; the maximum range is found at $\mu=0$. A
longer range can be obtained by interpolating
\begin{equation}
\alpha_{II}(x)=\alpha_0\ \left[
\ln\left(\frac{x_0+x}{x}\right) + \frac{\alpha_0}{\alpha_{sat}}
\right]^{-1}.
\label{qq35}
\end{equation}
We use $\Lambda_{QCD} = 0.2 \ \mbox{GeV}$ and $n_F=3$. The
dependence on $\Lambda_{QCD}$ and $n_F$, which is not large, can be
compensated for by modifying $\mu$ and $\alpha_{sat}$.

In the BSE the solutions are independent of the choice of gauge for
the vector part of the potential. This independence depends
essentially on the presence of crossed diagrams in the interaction
kernel~\cite{feldman}. However, in the BSLT and ET approximations
the relative energy is fixed, so crossed diagrams do not occur and
the gauge-independence of these quasi-potential equations must be
broken. In order to estimate the gauge-dependence of our results we
study two gauges. The first one is the Feynman gauge as used in
Eq.~(\ref{qq7}). The other is the Coulomb, transverse or radiation
gauge which is obtained by replacing
\begin{equation}
\gamma_\mu^{(1)} \gamma^{\mu (2)} \rightarrow
\gamma_\mu^{(1)} \gamma^{\mu (2)}
+\frac{1}{2} \left[
\mbox{\boldmath$\gamma$}^{(1)} \cdot\mbox{\boldmath$\gamma$}^{(2)}
-\frac{(\mbox{\boldmath$\gamma$}^{(1)} \cdot
\mbox{\boldmath$x$})(\mbox{\boldmath$\gamma$}^{(2)} \cdot
\mbox{\boldmath$x$})}{x^2}
\right]
\end{equation}
in Eq.~(\ref{qq7}). Appendix~\ref{appa3} gives the partial wave
projections of these potentials.

The presence of the $1/x$ term in the potential has important
consequences for the short distance behavior of the wave functions.
This can be understood from the following simple picture. Consider
the probability of the meson decaying through the annihilation of
the quark and the antiquark. This process is proportional to the
wave function at zero relative distance, or equivalently, to the
wave function integrated over all relative momenta. Using the wave
equation the integral over the wave function can be expressed as a
$\int dp\psi(p) = -\int dp S(p) \int dq V(p-q)\psi(q)$, that is, a
loop integral over the propagator $S$ and potential $V$ plus
additional corrections. This is illustrated diagrammatically in
Fig.~\ref{fig1}. Let us focus on the ultra-violet (uv) behavior of
the momentum integration over the triangle diagram. It is divergent
for most dynamical models due to the OGE potential. For example, in
the BSE the two-fermion propagator and the OGE interaction both fall
off as $p^{-2}$ for large $p$ (we neglect for the moment the
additional $\ln p^2$ behavior of the running coupling constant). So
for large relative momenta the loop integral takes the uv-divergent
form $\int\! d^4p\; p^{-4}$. Similarly, in the one-body Dirac
equation and in the two-body quasi-potential equations the
propagators behave as $p^{-1}$ and the potential as $p^{-2}$ what
also leads to a uv-divergent loop $\int\! d^3p\; p^{-3}$.
Divergences also occur in the light-cone formalism where the
propagator and potential go as $p_\bot^{-2}$ and the two quark
spinors as $p_\bot^2$~\cite{lepage} resulting in $\int\! d^2p_\bot\;
p_\bot^{-2}$. This suggests that the solutions of these wave
equations are singular at the origin. Indeed, at short distances the
Dirac wave functions~\cite{merz} and the Bethe-Salpeter wave
functions~\cite{soper,bastai} behave like $\psi(x) \sim x^{\gamma},\
-1<\gamma<0$, where $\gamma$ is a decreasing function of the
coupling constant $\alpha$ of the interaction. At some maximum value
of $\alpha$ the exponent $\gamma$ becomes smaller than $-1$ and
physical acceptable solutions do no longer exist. These features are
also shared by the BSLT and ET equations.

The extent to which these singularities actually appear in the
physical wave functions is weakened, if the fixed coupling constant
is replaced by a running coupling constant such that $\alpha(p) \sim
\ln^{-1} p^2$ for $p \rightarrow \infty$. The triangle loop,
however, is still uv divergent but the short distance
behavior is like $\psi(x) \sim \ln^{\gamma} |x|$. Furthermore,  the
singular behavior can be removed by renormalizing the wave function
by means of some cut-off scale. For example, this has been done in
the light-cone calculations of Ref.~\cite{lepage}. Yet, the
mathematical implications are important. Seemingly no attention has
been given to the asymptotic behavior in most work on relativistic
quark-quark dynamics such as in Refs.~\cite{jacob,lagae}. In
appendix B we analyze in detail the singular behavior of the wave
functions of the $J^P\!=\!0^-$ states (e.g. the pion) in the BSLT
equation with a fixed coupling constant. We have already discussed
the short distance behavior for the ET equation in
Ref.~\cite{peter}.

\section{PARAMETER DEPENDENCE}

In this section we describe the calculational procedure to solve the
quasi-potential equations and we present some results on the
parameter dependence of the spectrum of $J^{PC}\!=\!0^{-+}$ states.
For convenience we refer to these states as the $^1\!S_0$ states
since their wave functions contain mostly $^1\!S_0$ components.
Similarly, other $J^{PC}$ states are also named by their main
$^{2S+1}L_J$ components. Numerically stable solutions are obtained
by taking explicitly into account the singularity in the bound state
wave equation at $x=0$ due to the presence of the OGE term.  In
appendix B the  behavior of the wave function of the $^1\!S_0$ state
is analyzed in detail for small $x$. We find that the most singular
component of it behaves as $\psi \sim x^\gamma$, $-1<\gamma<0$, and
$\gamma$ given by Eqs.~(\ref{qq32}),~(\ref{qq33}), (\ref{qq36})
or~(\ref{qq34}). Let us introduce a function $f_\gamma(x)$ which for
small arguments behaves as $x^\gamma$ and for large arguments
becomes unity. By substituting $\psi(x) = f_\gamma(x)
\varphi(x)$ we get an equation for $\varphi(x)$ which is regular at
$x=0$. For the non $^1\!S_0$ states, no special precautions have to
be taken, because the wave function for the coupling constant
strengths considered in this work vanishes sufficiently fast in the
origin. Numerically accurate solutions are found by using the same
$x^\gamma$ factor as for $^1\!S_0$ channel. The same  applies when
we have a running  coupling constant, since the equations in this
case are less singular than those with a fixed coupling constant.

The partial wave projection of the BSLT and ET equations leads to
the coupled set of differential-integral equations~(\ref{qq20}) and
(\ref{qq21}) for $n$ ($3\!\leq\! n \!\leq\! 8$) components
$\psi_i(x)$, which are expanded as $\psi_i(x) = f_\gamma(x)
\sum_{j=1}^k c_{ij} S_j(x)$. The $S_j(x)$ are cubic Hermite spline
functions (see e.g. Ref.~\cite{payne}). The $\psi_i(x)$ are cut off
at some maximum value $x_{max}$ and we impose $\psi_i(x_{max})=
\psi_i'(x_{max})=0$. Near $x=0$ $\psi_i(x)$ is forced to be of order
$x^\gamma$ or higher.
By evaluating the equations~(\ref{qq20}) or (\ref{qq21}) at $k$
fixed points $x_1, ... ,x_k$ one obtains a set of $k \times n$
linear equations for the $k \times n$ spline coefficients  $c_{ij}$,
which only admits non-trivial solutions at the bound state energies.
It should be noted that due to the multiplication of the BSLT
equation by the operator $D$ of Eq.~(\ref{qq25}) a continuum of
additional solutions have been introduced in Eq.~(\ref{qq26}) for
$M>m_1+m_2$, which are not present in the original BSLT equation.
The cut-off makes these continuum unphysical solutions into a
discrete set, thus rendering it possible to isolate and reject them.
This is illustrated in Figs.~\ref{fig2}a and b which show two
$^1\!S_0$ solutions of the BSLT equation with $m=0.2\ \mbox{GeV},\
\kappa=0.2\ \mbox{GeV}^2,\
\varepsilon=\alpha=c=0$. The wave function shown in \ref{fig2}a is
the ground
state found at $M=1.23\ \mbox{GeV}$; the solution shown in
\ref{fig2}b found at $M=1.30\ \mbox{GeV}$ can easily be be
identified as an
outgoing wave. As a result the latter solution should be rejected.
The masses found for the physical bound states are insensitive to
variations of $x_{max}$, whereas the masses of the continuum
solutions do depend on $x_{max}$. Typical values that we used are
$x_{max} = 2\;\mbox{fm}$ and $k=30$ for the heavy quark systems up
to $x_{max} = 5\;\mbox{fm}$ and $k=120$ for the pion system.
For the latter system a
large number of splines is needed to obtain good accuracy since its
wave function can have a long oscillating tail. As an overall check
on the method the configuration space program was also used to
calculate the spectrum for a non-confining potential containing
scalar and vector exchanges. The resulting masses were verified by
solving the BSLT equation in momentum space \cite{fleischer};
agreement was found within $0.1\%$.

Let us now consider some important differences between the
non-relativistic and the relativistic equations. In the conventional
non-relativistic models as discussed for example in Ref.
\cite{isgur} the hyperfine interaction resulting from the OGE
contribution is singular near the origin and as result it has
to be regularized by introducing a phenomenological cutoff.  On the
other hand, our relativistic equations are well defined, at least up
to a critical coupling constant of the OGE interaction.  In this
case a natural cutoff essentially occurs due to relativity,  where
the scale is given by the mass of the constituent  quarks.  The
singular behavior of the wave function at $x=0$ induced by the
Coulomb-like interaction in the relativistic  equations considerably
modifies the meson mass if the the coupling constant $\alpha$ is
close to its maximum value. Fig.~\ref{fig3} shows the masses of a
mostly non-relativistic $^1\!S_0$ system ($m=5\ \mbox{GeV}$ and
$\kappa=0.2\ \mbox{GeV}^2$) as a function of the coupling constant.
For small $\alpha$ the spectrum agrees well with the Schr\"odinger
result. However, if the limit $\alpha \rightarrow \alpha_{max}$ is
taken, the relativistic results differ considerably;
$dM(\alpha)/d\alpha \rightarrow -\infty$ and $M(\alpha) \rightarrow
M_0$, where generally $M_0$ is not equal to zero.

Furthermore, for the non-relativistic Schr\"odinger-like equations
an additional constant $c$ in the potential causes solely an overall
shift in the meson mass spectrum. This does not happen in the
relativistic case. Because the quasi-potential equations are
time-reversal invariant each solution for a meson at some positive
mass is accompanied by its anti-meson solution at the same negative
mass. If the constant $c$ is added to the potential and we let $c$
grow to negative values, the absolute mass of both solutions
decrease until both mesons become massless. At this $c$ the
solutions coincide and for more negative $c$ no bound state can be
found except for higher excitations. This can clearly be seen from
Fig.~\ref{fig4} where the $^1\!S_0$ spectrum is plotted as a
function of $c$ for the BSLT and ET equations.

The replacement of the non-relativistic kinetic energy $p^2/2m$ by
the relativistic expression $\sqrt{p^2+m^2}$ and the introduction of
the negative energy states greatly reduces the spacings between the
ground state solution and the successive excitations for small quark
masses. Fig.~\ref{fig5} shows the calculated $^1\!S_0$ mass spectrum
as a function of the quark mass with the confinement taken such that
$\kappa/m^2$ is fixed. This confinement gives for the Schr\"odinger
equation constant binding energies. The figure illustrates nicely
that for large masses the non-relativistic results are obtained,
whereas for small masses the level density becomes considerably
higher.

Finally, another aspect which is absent in non-relativistic models
are the solutions describing a heavy quark and a light quark-hole.
Their presence is a consequence of the requirement that the one-body
Dirac equation is obtained in the limit that one particle becomes
very heavy, and they can be interpretated in an analogous way as the
negative energy states in the single fermion Dirac hole theory.
Fig.~\ref{fig6} shows the $^1\!S_0$ mass spectrum for fixed mass of
the first quark and various masses of the second quark. At equal
masses the hole-quark states are not present, but as $m_2$ grows
they emerge from the zero total mass axis, and at large $m_2$ two
spectra symmetrical around $M=\pm m_2$ are generated.

\section{LONG DISTANCE BEHAVIOR AND REGGE-TRAJECTORIES}

Experimentally it is found that the masses of the light mesons lie
on a linear Regge-trajectory, that is, for large angular momenta $J$
the squares of the masses $M$ of the mesons are proportional to
their angular momenta
\begin{equation}
M^2 = \beta J + c,
\end{equation}
where the Regge-slope $\beta \simeq 1.2 \
\mbox{GeV}^2$~\cite{lucha}. The value of the slope depends almost
exclusively on the confinement
strength and can be used to fix it. Figs.~\ref{fig7} and~\ref{fig8}
show the Regge behavior as obtained from the numerical solutions of
the BSLT and ET equations using $m_q=0.25\ \mbox{GeV},\
\alpha=0, c=-1.0\ \mbox{GeV}$ and a linear scalar confining
potential
with $\kappa=0.33\ \mbox{GeV}^2$. The figures indeed show a linear
relation between $M^2$ and $J$. However, the Regge-slope is
considerably smaller than predicted by non-relativistic models with
energy operator $\sqrt{p^2+m^2}$, where
$\beta=8\kappa$~\cite{lucha}; these models give a good description
of the c\=c and b\=b systems when $\kappa \simeq 0.18 \
\mbox{GeV}^2$ (see e.g.~\cite{isgur}). It turns out that a small
admixture of vector confinement $\varepsilon$, as in
Eq.~(\ref{qq7}), greatly affects the Regge-slopes. Figs.~\ref{fig7}
and~\ref{fig8} also show the Regge behavior for a fraction
$\varepsilon=0.15$ of vector confinement in the Feynman gauge and in
the Coulomb gauge. All trajectories are increased except for the
spin-singlet trajectory in the ET model in the Feynman gauge which
is decreased. Table~\ref{table1} summarizes the resulting
Regge-slopes.

A vector admixture in the confining interaction introduces a spin
dependence through the presence of the $|\!-+\rangle,\ |\!+-\rangle$
and $|\!--\rangle$ components.  Because of this when too much vector
confinement is chosen, $\varepsilon >0.2$, some light mesons are no
longer bound, i.e. certain channels may become deconfined.
This can be used to set an upper bound on the degree of allowed
vector admixture. Consider the strength of the confining potential
in the $^1J_J^e$ channel in the Feynman gauge. From Eq.~(\ref{qq18})
we see that the projection of $\gamma_\mu^{(1)} \gamma^{\mu (2)}$ in
this channel is $-4$, so the linear part of the q\=q-interaction in
this channel is
\begin{equation}
\kappa x  \left[ (1-\varepsilon) 1^{(1)} 1^{(2)}
+ \varepsilon \gamma_\mu^{(1)} \gamma^{\mu (2)}\right]
\stackrel{\textstyle\ \ \ ^1J_J^e\ \ \ }{\longrightarrow}
\kappa x  \left[ (1-\varepsilon)  -4 \varepsilon
\rule[-1.5mm]{0ex}{5mm}\right].
\end{equation}
Hence if $\varepsilon > 0.2$ the potential becomes repulsive for
large distances; the interaction tends to $-\infty$ for $ x
\rightarrow \infty$, which is clearly not physically admissionable.
This suggests that the $P=(\!-\!)^J$ mesons of the light quark
system, of which the wave functions contain the $^1J_J^e$ channel
and have $|\!--\rangle$ components of size comparable to the
$|\!++\rangle$ components, do not have a bound state solution if
$\varepsilon > 0.2$. This is confirmed by the numerical solutions.
In the Coulomb gauge a slightly larger value is allowed, namely
$\varepsilon = 0.25$. In order to avoid these problems we demand
that the admixture of the vector interaction in the confining
potential does not reach or exceed these values.

It is interesting to note that also in the one-body Dirac equation,
which can be used to describe a system of a very heavy and a light
quark a similar problem arise. Such a framework has recently been
used to study the spectrum of the $D$ and $B$ mesons~\cite{simonov}.
A pure scalar confinement interaction was used. The repulsion
exercised by the vector confinement on the negative components gives
also in this case an upper limit on the  allowed degree of vector
admixture in the confining potential. Consider the Dirac equation
\begin{equation}
(m-i\!\not\! \partial)\psi(\mbox{\boldmath$x$}) =
-\kappa x [ (1-\varepsilon) + \varepsilon \gamma_0 ]
\psi(\mbox{\boldmath$x$}).
\end{equation}
Projection upon states of definite $J$ gives
\begin{equation}
\left( \begin{array}{cc}
m-E & D_{J+3/2} \\ D_{-J+1/2} & m+E
\end{array} \right)
\psi(x) = -\kappa x
\left( \begin{array}{cc}
1 & 0\\
0 & 1-2\varepsilon
\end{array} \right)
\psi(x),
\end{equation}
and $D_\ell$ as defined in Eq.~(\ref{qq11}). Eliminating the lower
component of $\psi(x)$ in favor of the upper component of
$\psi_1(x)=F(x)/x$ and letting $x \rightarrow \infty$ leads to
\begin{equation}
\left\{ \left[
-\frac{\partial^2 \hfill}{\partial x^2}
+ \frac{1}{x}\frac{\partial \hfill}{\partial x}
+ \frac{1}{x^2} (J-3/2)(J+1/2) \right]
+ (1-2\varepsilon)\kappa^2 x^2 \right\} F(x) = 0.
\end{equation}
For $\varepsilon > 0.5 $ the potential term becomes negative and the
solutions for $F(x)$ are no longer bound.

\section{MESON MASS SPECTRUM}

In Table~\ref{table2} the parameters are listed that we used to
calculate the meson mass spectrum. We did not include the light
isoscalar mesons ($\eta,\ f,\ h$). The masses of these mesons are
expected to be modified by the process of q\=q annihilation and
creation, making the transitions
u\=u$\leftrightarrow$d\=d$\leftrightarrow$s\=s likely.

We determine the strength of the confinement from the condition that
the experimental Regge-slope $\beta\simeq 1.2\ \mbox{GeV}^2$ is
found. If the confinement is taken purely scalar then a confinement
is needed which is much stronger than the strength known from
non-relativistic models to give a good description of the c\=c and
b\=b systems. Therefore the maximum of allowed vector admixture is
chosen which gives a maximum Regge-slope at a fixed confinement
strength. So in the BSLT model we take $\varepsilon=0.2$ and
$\varepsilon=0.25$ in the Feynman gauge and in the Coulomb gauge,
respectively. With these values the experimental Regge-slopes are
found at $\kappa\simeq0.33\ \mbox{GeV}^2$. Although this strong
confinement makes it difficult to get a perfect fit for the heavy
quark systems we need it in order to get an acceptable light meson
description. As can be seen from Table~\ref{table1} and
Figs.~\ref{fig7} and~\ref{fig8}, an increase of vector confinement
in the Feynman gauge in the ET model gives an increase in the
Regge-slopes at the spin-triplet states but a decrease at the
spin-singlet states. It is therefore impossible in the ET model in
the Feynman gauge to obtain simultaneously a fair description of the
heavy mesons and of the Regge-slopes of the light mesons, so for the
fit of the meson mass spectrum we only used the ET model in the
Coulomb gauge. For this last case we took the value
$\varepsilon=0.2$ which is slightly less than the maximum
$\varepsilon=0.25$; above $\varepsilon \simeq 0.2$ the spatial
extension of the $^1J_J^e$ components becomes more than $\sim\!\!
10\ \mbox{fm}$ which is unphysically large.

For the OGE part of the potential the choice of the saturation value
$\alpha_{sat}$ of the running coupling constant is not important
since it only plays a role at large distances were the confining
part dominates. Fair fits can be found with $\alpha_{sat}=0.6$ as
well with $\alpha_{sat}=1.0$. We choose the commonly accepted value
$\alpha_{sat}=0.8$. The typical range of the running coupling
constant regime is much more important. In the Coulomb gauge the
maximum fixed coupling constants allowed in the BSLT and in ET model
are respectively $\alpha_{max}=8/(3\pi) \simeq 0.85$ and
$\alpha_{max}=4/(3\pi) \simeq 0.42$ [see Eqs.~(\ref{qq36}) and
(\ref{qq34})]. If the running coupling constant regime is small and
the saturation value is approximately equal or larger than the
maximum allowed fixed coupling constant, $\alpha_{sat}
\mbox{\raisebox{-0.6ex}{$\stackrel{>}{\sim}$}} \alpha_{max}$, then
at medium small distances the wave function exhibits the singular
behavior of the fixed coupling constant equations. Especially, the
spectrum shows much lower energies for the $^1\!S_0$ states than the
non-relativistic spectrum. Since $\alpha_{max}$ is rather low in the
ET model this effect occurs rather strongly in the ET spectrum, and
one must choose a much larger running coupling regime for the ET
equations than for the BSLT equations. So we use for the BSLT model
the coupling $\alpha_I$ of Eq.~(\ref{qq27}) and for the ET model the
coupling $\alpha_{II}$ of Eq.~(\ref{qq35}). Yet, even with this
choice the $^1\!S_0$ states are considerably lower in the ET fit
than in the BSLT fits.

The masses of the u and d quarks are taken equal and chosen together
with the constant $c$ to give a fair description of the light
non-strange mesons. Next the masses of the s, c, and b quarks are
fitted to the $^3\!S_1$ states $K^*$, $J/\psi$ and $\Upsilon$,
except for the c mass in the ET model where a correct $K^*$ mass
would lead to a far too low $K$ mass.

Table~\ref{table3} presents the resulting mass spectra of the BSLT
approach in both the Feynman and in the Coulomb gauge and of the ET
approach in the Coulomb gauge, together with the known experimental
mass spectrum. As can be seen from these numbers the BSLT approach
in the Feynman gauge gives the best description, but the differences
between the BSLT results obtained from the Feynman gauge and those
from the Coulomb gauge are minor. The ET model does not give a very
good spectrum.

Let us discuss the spectra in more detail. The ground states of the
d\=d and d\=s systems, the $\pi$ and $K$ mesons, are considerably
lighter than the BSLT fits predict. The light masses of these mesons
are commonly explained within in the framework of broken chiral
symmetry where these mesons correspond to almost massless Goldstone
bosons. Since the quasi-potential equations do not incorporate the
chiral symmetry they give too heavy $\pi$ and $K$. The correct mass
of the $\pi$ in the ET is a coincidence and due to the too singular
behavior of the $^1\!S_0$ states in the ET model.

The description of the $^1\!P_1$-states in the ET model is not good.
According to the Breit-reduction of the potential, the mass of this
state should be equal to the center of gravity of the $^3\!P$ states
up to order $1/m_q^2$. Clearly none of the $^1\!P_1$-states in the
ET model satisfy this condition, in contrast to the BSLT model where
this condition is more or less satisfied. Furthermore the splittings
between the various $^3\!S_1$ and $^3\!D_1$ states, especially in
the d\=d and d\=s system, are much too small in the ET model. We
conclude that the ET model does not give the correct fine structure.
At this point we would like to note that within a constituent quark
model with flavor independent potential it is not possible to have
simultaneously a correct description of the $P$-states in the d\=d
and d\=s systems. The only difference between these systems is that
one of the constituent quarks is a little heavier, so it can be
expected that the levels of the d\=s system are raised a little as
compared to the levels of the d\=d system, with a little smaller
fine-structure splitting. But the observed spacings between the
$P$-states of the d\=d and d\=s mesons do not follow this pattern.

Considerable deviations between the experimental masses and the
calculated masses appear in the excited $P$-states of the b\=b
system. This is due to the large confinement strength $\kappa$ that
was chosen in order to have proper Regge-slopes for the light
mesons. We have searched for various simple prescriptions to improve
on the shape of the potential in such a way that the masses of the
excited $P$-states of the bottonium system show better agreement.
The large confining potential needed to get the proper Regge-slope
for the light mesons and the weaker confinement needed for the heavy
excited $P$-states in bottonium clearly suggest that maybe a better
overall description can be found if one takes a smaller $\kappa$ at
short distances than at large distances. We examined this
possibility by modifying the potential at short distances according
to
\begin{equation}
V_{conf}(x) \rightarrow \tilde{V}_{conf}(x) = \left\{
\begin{array}{ll}
\displaystyle
\kappa' x + c'+Ax^2\ \ \ \ & x <x_0, \\
\displaystyle
\kappa x + c & x >x_0,\rule{0ex}{3ex}
\end{array}\right.
\end{equation}
with the two regions smoothly matching, $\kappa' < \kappa$, $A>0$
and various character (vector or scalar) for
$\tilde{V}_{conf}-V_{conf}$. We found that it was indeed possible to
get the right Regge-slopes for the light mesons and simultaneously
the b\=b excitations at the correct levels. But this also increased
the fine-structure splittings in the non-strange $P$-states up to
values as $0.3 \ \mbox{GeV}$ for the difference $M(a_2)-M(a_1)$
which is known to be only 0.051 \ \mbox{GeV}. Furthermore, the
decrease of the splitting $M(2^3P_1)-M(1^3S_1)$ in b\=b causes
almost a similar reduction of the shift $M(1^3P_1)-M(1^3S_1)$ in
c\=c which is undesirable. These two differences depend more or less
in the same way on the potential since the excited levels $2^3P_1$
of b\=b and $1^3P_1$ of c\=c both have approximately the same
spatial extension.

\section{CONCLUDING REMARKS}

We studied two approximations to the Bethe-Salpeter equation for the
two quark system, the Blankenbecler-Sugar-Logunov-Tavkhelidze (BSLT)
equation and an equal-time (ET) equation. In these quasi-potential
approximations the relative energy dependence of the wave function
was eliminated by assuming a simplified form for the two-body
propagator. The full Dirac structure of positive and negative energy
states was kept. Both two-body propagators reduce to the one-body
Dirac propagator if one of the particles is taken infinitely heavy.
We applied these equations to a system of two constituent quarks
interacting through a phenomenological potential which consisted of
two parts. The first part was a Coulomb-like one-gluon exchange
(OGE) part, the second part was a linear confining potential which
was taken mostly scalar-like and partly vector-like. For the vector
part we studied both the Feynman and the Coulomb gauge. Since the
confining potential is highly singular in momentum space we
transformed the quasi-potential equations to configuration space.

It was shown that for a fixed coupling constant for the OGE
potential a maximum value exists ---depending on the model and
gauge--- above which the ground state solutions no longer exist.
Also the admixture of vector character in the confining potential
has a maximum value (20\% and 25\% in the Feynman and Coulomb gauge
respectively) above which some mesons become unbound. The latter
could be explained from the repulsion between positive and negative
energy states. We found linear Regge-trajectories for both models
and gauges, but their Regge-slopes were much smaller than predicted
by models with only positive energy states.

Using the BSLT and ET equations we calculated the full known meson
mass spectrum of all light and heavy mesons.  As compared to the
non-relativistic model predictions \cite{isgur} given the limited
number of parameters used here, the fit can be considered
satisfactory. We believe that our prediction of the spectrum is of
comparable or better quality than other relativistic studies based
on quasi-potential and Dirac equations \cite{mitra,crater}. We found
only a small gauge dependence in the BSLT spectrum. For the ET
equation we only used the Coulomb gauge since it was impossible to
get satisfying Regge-slopes in the Feynman gauge. The fine-structure
of the spectrum, such as the spacings between the $P$-states and
between the $^3\!S_1$ and $^3\!D_1$ states, as calculated from the
ET equations did not follow the experimental spacings nor did it
follow predictions of the Breit-reduction of the potential. The BSLT
fine-structure showed more or less agreement with these. However,
from the agreement that was found between the calculated meson mass
spectrum and the experimental spectrum we conclude that for a
calculation able to reproduce all meson masses within $\sim 0.03\
\mbox{GeV}$ one cannot suffice with a constituent quark model and a
phenomenological flavor independent
potential. Specifically, one would like to incorporate chiral
symmetry in order to reproduce the correct $\pi$ and $K$ mass and
some flavor dependence in the confinement strength in order to
reduce this strength in the heavy quark systems.

The wave functions found in these calculations can in principle be
used to perform relativistic covariant calculations on interactions
between mesons and other particles. Especially at high momenta
significant deviations from the non-relativistic calculations can be
expected; one example has been discussed in Ref.~\cite{peter}.

\acknowledgments
This work was partially financially supported by de Stichting voor
Fundamenteel Onderzoek der Materie (FOM), which is sponsored by the
Nederlandse Organisatie voor Wetenschappelijk Onderzoek (NWO).

\appendix{PARTIAL WAVE PROJECTION}

In this appendix we describe the projection of the potential and the
BSLT and ET equations on sets of basis states with definite parity
and total angular momentum. The equations and basis states are
formulated for a two quark system; they can easily be transformed to
a quark-antiquark system by performing a charge-conjugation on one
of the quarks. The complete meson wave function is written as a
combination of the states $| JLSJ_z\!\!>\! \otimes |\rho_z^{(1)}
\rho_z^{(2)} \!\!>\! \otimes \chi_{color} \otimes \chi_{flavor}$.
The color and flavor parts of the wave function are well-known and
can be found e.g. in Ref.~\cite{isgur}. For the $\rho$-spin basis
states we do not use the energy eigenstates projected out by the
$\Lambda^{\rho_1 \rho_2}$, but instead the four eigenstates of
$\gamma_0^{(1)} \gamma_0^{(2)}$, corresponding to the four
combinations of upper and lower components of the quark spinors. The
energy eigenstates have the advantage that the propagator is simply
diagonal on this basis as is shown by Eqs.~(\ref{qq4}) and
(\ref{qq14}). However, on this basis the matrix elements of the
potential are complicated and many in number. In principle it is
possible to transform them all to configuration space, but the
numerical work involved is huge. On the other hand, on the basis of
the eigenstates of $\gamma_0^{(1)} \gamma_0^{(2)}$ the matrix
elements of the potential are almost trivial, whereas the projection
of the propagator on this set gives only a limited number of
functions to be transformed. For the BSLT propagator there is only
one such function namely $Z_{BSLT}$~(\ref{qq5}), whereas for the ET
propagator two functions $Z_1$ and $Z_2$~(\ref{qq15}) are needed.

\subsection{Blankenbecler-Sugar-Logunov-Tavkhelidze equation}
\label{appa1}

We decompose the kinetic part of the BSLT propagator into pieces
acting in the subspaces of the spin and rho-spin:
\begin{eqnarray}
\not\!{\tilde p}_1 + \not\!{\tilde p}_2 & = &
 i\mbox{\boldmath$\gamma$}^{(1)} \cdot \mbox{\boldmath$\nabla$} +
E_1 \gamma_0^{(1)}
-i\mbox{\boldmath$\gamma$}^{(2)} \cdot \mbox{\boldmath$\nabla$} +
E_2 \gamma_0^{(2)}
\nonumber \rule[-4mm]{0ex}{4mm} \\
&= &
-\rho_2^{(1)} \mbox{\boldmath$\sigma$}^{(1)} \cdot
\mbox{\boldmath$\nabla$} + E_1 \rho_3^{(1)}
+\rho_2^{(2)} \mbox{\boldmath$\sigma$}^{(2)} \cdot
\mbox{\boldmath$\nabla$} + E_2 \rho_3^{(2)}.
\end{eqnarray}
For a given total spin $J$ there are four basis states for the spin
subspace. In the spectroscopic notation $^{2S+1}L_J$ they are
\{$ ^1J_J$, $^3J_J$, $^3(J\!-\!1)_J$, $^3(J\!+\!1)_J $\}.
After some Clebsch-Gordan algebra one finds that on this basis
\begin{equation}
\mbox{\boldmath$\sigma$}^{(1)} \cdot \mbox{\boldmath$\nabla$} =
\left( \begin{array}{cccc}
0&0& c_1 D_{-J+1} &-c_2 D_{J+2} \\
0&0&-c_2 D_{-J+1}&-c_1 D_{J+2} \\ c_1 D_{J+1}&-c_2 D_{J+1}&0&0 \\
-c_2 D_{-J}&-c_1 D_{-J}&0&0
\end{array} \right),
\end{equation}
where
\begin{equation}
D_\ell = \frac{\partial\hfill}{\partial x} + \frac{\ell}{x}, \ \ \ \
c_1 = \sqrt{\frac{J}{2J+1}}, \ \ \ \
c_2 = \sqrt{\frac{J+1}{2J+1}}.
\label{qq11}
\end{equation}
The matrix of $\mbox{\boldmath$\sigma$}^{(2)} \cdot
\mbox{\boldmath$\nabla$}$ is identical, except for the first row and
first column ---these correspond to the antisymmetric spin singlet
state--- which change sign. Let us combine the four eigenstates of
$\gamma_0^{(1)}
\gamma_0^{(2)}$ into the rho-spin combinations
\begin{equation}
s  =  \frac{|\!+\!+\rangle\! - |\!--\rangle}{\sqrt{2}},\ \  a  =
\frac{|\!+\!+\rangle\! + |\!--\rangle}{\sqrt{2}},\ \  e  =
\frac{|\!+\!-\rangle\! + |\!-+\rangle}{\sqrt{2}},\ \
o  =  \frac{|\!+\!-\rangle\! - |\!-+\rangle}{\sqrt{2}}.
\label{qq8}
\end{equation}
Phase factors $i$ between the eigenstates are used according to
\begin{equation}
|\!+\!+\rangle =
-i\rho_1^{(1)} |\!-+\rangle =
-i\rho_1^{(2)} |\!+-\rangle =
\rho_1^{(1)} \rho_1^{(2)} |\!--\rangle.
\end{equation}
This ensures that the final equations will be real. Following Kubis
{}~\cite{kubis} and Gammel~\cite{gammel} we combine the spin and
rho-spin bases to form the basis set ${\cal B}$ of eight states with
unnatural parity $P=(\!-\!)^{J+1}$
\begin{equation}
{\cal B} = \{ ^1J^s_J,\ ^1J^a_J,\ ^3J^s_J,\ ^3J^a_J,\
^3(J\!-\!1)^e_J,\ ^3(J\!-\!1)^o_J,\ ^3(J\!+\!1)^e_J,\
^3(J\!+\!1)^o_J\},
\label{qq9}
\end{equation}
and the basis set ${\cal B}^*$ of eight states with natural parity
$P=(\!-\!)^J$
\begin{equation}
{\cal B}^* =\{ ^1J^e_J,\ ^1J^o_J,\ ^3J^e_J,\ ^3J^o_J,\
^3(J\!-\!1)^s_J,\ ^3(J\!-\!1)^a_J,\ ^3(J\!+\!1)^s_J,\
^3(J\!+\!1)^a_J\}.
\label{qq10}
\end{equation}
These basis states for two quarks can be related to the often used
Dirac matrix set for a quark and an anti-quark~\cite{llewell2} by
performing a charge conjugation on one of the spinors. The relation
is shown in Table~\ref{table4}. Let us expand the expression for
$\not\!{\tilde p}_1 + \not\!{\tilde p}_2$ on the set ${\cal B}$
(\ref{qq9}) of unnatural parities and denote the resulting
$8\times8$ matrix by $\mbox{\boldmath$D$}(x)$. This gives
\begin{equation}
\frac{\mbox{\boldmath$D$}}{2}=
\left( \begin{array}{cccccccc}
 0 &E & 0 & 0 &-c_1D_{-J+1}& 0 &c_2D_{J+2}& 0 \\ E & 0 & 0 & 0 & 0 &
0
& 0 & 0 \\
 0 & 0 & 0 &E & 0 & 0 & 0 & 0 \\
 0 & 0 &E & 0 & 0 &-c_2D_{-J+1}& 0 &-c_1D_{J+2}\\
-c_1D_{J+1}& 0 & 0 & 0 & 0 &\Delta & 0 & 0 \\
 0 & 0 & 0 &-c_2D_{J+1}&\Delta & 0 & 0 & 0 \\  c_2D_{-J}& 0 & 0 & 0
&
0 & 0 & 0 &\Delta \\
 0 & 0 & 0 &-c_1D_{-J}& 0 & 0 &\Delta & 0
\end{array} \right),
\end{equation}
with $2E=E_1+E_2=M$ and $2\Delta=E_1-E_2=(m_1^2 - m_2^2)/M$. The
matrix $\mbox{\boldmath$D$}^*$ of $\not\!{\tilde p}_1 +
\not\!{\tilde p}_2$ for the set ${\cal B}^*$~(\ref{qq10}) of natural
parity states can be obtained from
\mbox{\boldmath$D$} by interchanging $E \leftrightarrow \Delta$.
The
complete partial wave projected BSLT equation is the following set
of eight coupled integro-differential equations
\begin{equation}
[\mbox{\boldmath$D$}(x) - \mbox{\boldmath$M$}]
\mbox{\boldmath$\psi$}(x) =
\frac{1}{2(m_1+m_2)} \int \! dx' \; x'^2\
\mbox{\boldmath$Z$}_{BSLT}(x,x')
[\mbox{\boldmath$D$}(x') + \mbox{\boldmath$M$}]
\mbox{\boldmath$V$}(x') \mbox{\boldmath$\psi$}(x').
\label{qq20}
\end{equation}
The matrix \mbox{\boldmath$M$} is equal to $m_1+m_2$ times the
identity, and the diagonal matrix $\mbox{\boldmath$Z$}_{BSLT}$ has
as i-th diagonal element the projection of $Z_{BSLT}$ on the orbital
angular momentum $\ell_i$ of the i-th basis state,
\begin{equation}
Z_{BSLT}^{(\ell)}(x,x')= 2\pi \int_{-1}^1 \!\!d\gamma\;
P_\ell(\gamma)
\,Z_{BSLT}(\sqrt{x^2+x'^2-2xx'\gamma}).
\end{equation}
The number of coupled channels becomes less if the quarks have equal
masses. In that case $\Delta = 0$ and the matrix \mbox{\boldmath$D$}
for the the unnatural parity $P=(\!-\!)^{J+1}$ states falls apart
into two $4\times 4$ parts describing the charge-parity
$C=(\!-\!)^J$ states and the $C=(\!-\!)^{J+1}$ states
\begin{equation}
\{ ^1J^s_J,\ ^1J^a_J,\ ^3(J\!-\!1)^e_J,\ ^3(J\!+\!1)^e_J \}
\ \ \ \mbox{and}\ \ \
\{ ^3J^s_J,\ ^3J^a_J,\ ^3(J\!-\!1)^o_J,\ ^3(J\!+\!1)^o_J\},
\label{qq19}
\end{equation}
respectively. If the masses are equal the set of natural parity
$P=(\!-\!)^J$ states can be split into the $C=(\!-\!)^J$ states and
the $C=(\!-\!)^{J+1}$ states
\begin{equation}
\{ ^1J^e_J,\ ^3J^o_J,\ ^3(J\!-\!1)^s_J,\ ^3(J\!-\!1)^a_J,
\ ^3(J\!+\!1)^s_J,\ ^3(J\!+\!1)^a_J\}
\ \ \ \mbox{and}\ \ \
\{ ^1J^o_J,\ ^3J^e_J\},
\end{equation}
respectively. The latter $C=(\!-\!)^{J+1}$ set contains no
$|\!++\rangle$ components and therefore it has no non-relativistic
analog. However since Eq.~(\ref{qq20}) is independent of E for this
case, no bound states exist in this sector.

\subsection{Equal-time approximation}
\label{appa2}

As with the BSLT propagator, the parts ${{\cal
S}}_1(\mbox{\boldmath$x$})$ and ${{\cal S}}_2(\mbox{\boldmath$x$})$
in Eqs.~(\ref{qq16}) and~(\ref{qq17}) of the ET propagator can be
projected on the partial wave bases. One finds for the set ${\cal
B}$ (\ref{qq9}) of unnatural parity states
\begin{eqnarray}
&&\mbox{\boldmath$S$}_1= m_1m_2\mbox{\boldmath$I$} +
\rule[-4mm]{4mm}{0ex}
\nonumber \\
&&
\left( \begin{array}{cccccccc}
\partial^2_J & -m_1E_2 &
   0 & 0 &
   c_1AD_{-\!J\!+\!1} & c_1E_2D_{-\!J\!+\!1} &
   -c_2AD_{J\!+\!2} & -c_2E_2D_{J\!+\!2} \\
-m_1E_2 &-\partial^2_J &
   0 & 0 &
   -c_1E_2D_{-\!J\!+\!1} &c_1BD_{-\!J\!+\!1} &
   c_2E_2D_{J\!+\!2}  &-c_2BD_{J\!+\!2} \\
0 & 0 &
   -\partial^2_J & -m_1E_2 &
   c_2BD_{-\!J\!+\!1} & -c_2E_2D_{-\!J\!+\!1} &
   c_1BD_{J\!+\!2} & -c_1E_2D_{J\!+\!2} \\
0 & 0 &
   -m_1E_2 &\partial^2_J &
   -c_2E_2D_{-\!J\!+\!1} &c_2AD_{-\!J\!+\!1} &
   c_1E_2D_{J\!+\!2}  &c_1AD_{J\!+\!2} \\
   c_1AD_{J\!+\!1} & -c_1E_2D_{J\!+\!1} &
   c_2BD_{J\!+\!1} & c_2E_2D_{J\!+\!1} &
   -\frac{1}{2J\!+\!1}\partial^2_{J-1} & m_1E_2 &
   -2c_1c_2H_2 & 0\\ c_1E_2D_{J\!+\!1} & c_1BD_{J\!+\!1} &
   -c_2E_2D_{J\!+\!1}& c_2AD_{J\!+\!1} &
   m_1E_2 &\frac{1}{2J\!+\!1}\partial^2_{J-1}  &
   0 &2c_1c_2H_2 \\
-c_2AD_{-\!J} & c_2E_2D_{-\!J} &
   c_1BD_{-\!J} & c_1E_2D_{-\!J} &
   -2c_1c_2H_1 & 0 &
   \frac{1}{2J\!+\!1}\partial^2_{J\!+\!1} & m_1E_2 \\
-c_2E_2D_{-\!J} &-c_2BD_{-\!J} &
   -c_1E_2D_{-\!J} &c_1AD_{-\!J} &
   0 &2c_1c_2H_1 &
   m_1E_2&-\frac{1}{2J\!+\!1}\partial^2_{J\!+\!1}
\end{array} \right)
\nonumber \\
&&
\end{eqnarray}
with $D_\ell,\ c_1,\ c_2$ as defined in~(\ref{qq11}), $A=m_1+m_2,\
B=m_1-m_2$, and
\begin{eqnarray}
H_1&=&\frac{\partial^2\hfill}{\partial x^2}
+\frac{-2J+1}{x}\frac{\partial\hfill}{\partial
x}-\frac{-J^2+1}{2x^2},
\\
H_2&=&\frac{\partial^2\hfill}{\partial x^2}
+\frac{2J+3}{x}\frac{\partial\hfill}{\partial
x}-\frac{-J^2-2J}{2x^2},
\rule[-6.5mm]{0ex}{15mm}
\\
\partial_\ell^2&=&\frac{\partial^2\hfill}{\partial x^2}
+\frac{2}{x}\frac{\partial\hfill}{\partial
x}-\frac{\ell(\ell+1)}{2x^2}.
\end{eqnarray}
The projection $\mbox{\boldmath$S$}_1^*$ of ${\cal S}_1$ for the set
${\cal B}^*$~(\ref{qq10}) of natural parity states is given by
\begin{equation}
\mbox{\boldmath$S$}_1^* = \mbox{\boldmath$T$} \mbox{\boldmath$S$}_1
\mbox{\boldmath$T$},\ \ \
\mbox{\boldmath$T$}=\mbox{diag}(1,-1,1,-1,1,-1,1,-1).
\end{equation}
The projection $\mbox{\boldmath$S$}_2$ of ${\cal S}_2$ for the set
${\cal B}$ can be expressed as
\begin{equation}
\mbox{\boldmath$S$}_2(m_1,m_2,E_1,E_2) = \mbox{\boldmath$C$}
\mbox{\boldmath$S$}_1(m_2,m_1,E_2,E_1) \mbox{\boldmath$C$},
\end{equation}
with diagonal charge-parity matrix
$\mbox{\boldmath$C$}=\mbox{diag}(1,1,-1,-1,1,-1,1,-1)$. And finally,
\begin{equation}
\mbox{\boldmath$S$}_2^*=\mbox{\boldmath$T$} \mbox{\boldmath$S$}_2
\mbox{\boldmath$T$} \ \ \ \mbox{or}\ \ \
\mbox{\boldmath$S$}^*_2(m_1,m_2,E_1,E_2) = \mbox{\boldmath$C$}^*
\mbox{\boldmath$S$}^*_1(m_2,m_1,E_2,E_1) \mbox{\boldmath$C$}^*,
\end{equation}
with charge-parity matrix $\mbox{\boldmath$C$}^*=\mbox{\boldmath$T$}
\mbox{\boldmath$C$}=\mbox{diag}(1,-1,-1,1,1,1,1,1)$. The complete
partial wave projected ET equation becomes the following set of
eight coupled integro-differential equations
\begin{equation}
\int \! dx' \; x'^2
\left[\frac{1}{m_1} \mbox{\boldmath$Z$}_1(x,x')
\mbox{\boldmath$S$}_1(x') +
\frac{1}{m_2} \mbox{\boldmath$Z$}_2(x,x') \mbox{\boldmath$S$}_2(x')
\right]
\mbox{\boldmath$\psi$}(x') = - \mbox{\boldmath$V$}(x)
\mbox{\boldmath$\psi$}(x).
\label{qq21}
\end{equation}
The diagonal matrices $\mbox{\boldmath$Z$}_i$ have as j-th diagonal
element the projection of $Z_i$ on the orbital angular momentum
$\ell_j$ of the j-th basis state,
\begin{equation}
Z_i^{(\ell)}(x,x')= 2\pi \int_{-1}^1 \!\!d\gamma\; P_\ell(\gamma)
\,Z_i(\sqrt{x^2+x'^2-2xx'\gamma}).
\end{equation}

\subsection{Potential}
\label{appa3}

The vector structure of the q\=q-interaction gives
\begin{equation}
\gamma_\mu^{(1)} \gamma^{\mu (2)}=
\rho_3^{(1)}\rho_3^{(2)} + \rho_2^{(1)}\rho_2^{(2)}
\mbox{\boldmath$\sigma$}^{(1)} \cdot \mbox{\boldmath$\sigma$}^{(2)}
=
\left( \begin{array}{cccc}
1+f & 0 & 0 & 0 \\
0 & 1-f & 0 & 0 \\
0 & 0 &-1+f & 0 \\
0 & 0 & 0 &-1-f
\end{array} \right)
\label{qq18}
\end{equation}
on the basis~(\ref{qq8}). Here $f=-3$ for spin singlet states and
$f=1$ for spin triplet states.  Choosing the Coulomb gauge amounts
to replacing
\begin{equation}
\mbox{\boldmath$\sigma$}^{(1)} \cdot \mbox{\boldmath$\sigma$}^{(2)}
\rightarrow
\frac{1}{2}\left[
\mbox{\boldmath$\sigma$}^{(1)} \cdot \mbox{\boldmath$\sigma$}^{(2)}
+ \frac{(\mbox{\boldmath$\sigma$}^{(1)} \cdot
\mbox{\boldmath$x$})(\mbox{\boldmath$\sigma$}^{(2)} \cdot
\mbox{\boldmath$x$})}{x^2}
\right].
\end{equation}
Thus in the Coulomb gauge the spin-spin interaction is reduced. The
difference with the Feynman gauge can be expressed as
\begin{equation}
\frac{1}{2} \left[\mbox{\boldmath$\gamma$}^{(1)}
\cdot\mbox{\boldmath$\gamma$}^{(2)}
-\frac{(\mbox{\boldmath$\gamma$}^{(1)} \cdot
\mbox{\boldmath$x$})(\mbox{\boldmath$\gamma$}^{(2)} \cdot
\mbox{\boldmath$x$})}{x^2}
 \right] =
\left( \begin{array}{cccccccc}
1 & 0 & 0 & 0 & 0 & 0 & 0 & 0 \\
0 &-1 & 0 & 0 & 0 & 0 & 0 & 0 \\
0 & 0 & 0 & 0 & 0 & 0 & 0 & 0 \\
0 & 0 & 0 & 0 & 0 & 0 & 0 & 0 \\
0 & 0 & 0 & 0 &-c_1^2 & 0 &c_1c_2 & 0 \\
0 & 0 & 0 & 0 & 0 &c_1^2 & 0 & -c_1c_2 \\
0 & 0 & 0 & 0 &c_1c_2 & 0 &-c_2^2 & 0 \\
0 & 0 & 0 & 0 & 0 &-c_1c_2 & 0 &c_2^2
\label{qq22}
\end{array} \right)
\end{equation}
both for the bases~(\ref{qq9}) and~(\ref{qq10}). The constants $c_i$
have been given in Eq.~(\ref{qq11}).

\appendix{SHORT DISTANCE BEHAVIOR OF $^1\!S_0$ STATES}

Here we discuss in detail  the singular behavior of the wave
function of the $J^P=0^-$ states (e.g. pion), to which we refer to
as $^1\!S_0$ states for convenience because their wave functions
contain mostly $^1\!S_0$ components. At short distances, or
equivalently at high momenta, the masses of the two quarks can be
neglected, so the short distance behavior for the unequal quark mass
case is identical to the equal quark mass case. Thus in this
appendix only the equal mass case needs to be considered. Since the
short distance behavior for the ET equation has already been
discussed in Ref.~\cite{peter} we concentrate on the BSLT
approximation. Since $J=0$ the $^3(J-1)^e_J$ state is removed from
the basis in~(\ref{qq19}) and the wave function is decomposed as
\begin{equation}
\psi(x, \Omega) = \frac{1}{x} \left[
\psi_1(x) | ^1S^s_0 >  + \psi_2(x) | ^1S^a_0 > + \psi_3(x) | ^3P^e_0
>
\rule[-1.7ex]{0ex}{4.5ex}
\right].
\end{equation}
In this basis the BSLT equation~(\ref{qq20}) reduces to
\begin{displaymath}
\left(\begin{array}{ccc}
-m & E & D_1 \\ E& -m  & 0 \rule[-2ex]{0ex}{5ex}\\ D_{-1}& 0 & -m
\end{array}\right)
\left(\begin{array}{c} \psi_1(x) \\
\rule[-1.7ex]{0ex}{4.5ex}\psi_2(x) \\ \psi_3(x)
\end{array}\right)
=
\frac{1}{4m} \int_0^\infty\! dx'\ xx'
\left( \begin{array}{ccc}
  Z^{(0)}(x,x') & 0 & 0 \\
  0 & Z^{(0)}(x,x') & 0\rule[-1.7ex]{0ex}{4.5ex} \\
  0 & 0 & Z^{(1)}(x,x')
\end{array} \right)
\end{displaymath}
\begin{equation}
\times
\left(\begin{array}{ccc}
m & E & D_1 \\ E& m  & 0 \rule[-1.7ex]{0ex}{4.5ex}\\ D_{-1}& 0 & m
\end{array}\right)
\left[
\left( \begin{array}{ccc}
-2 & 0 & 0 \\
0 & 4 & 0 \rule[-1.7ex]{0ex}{4.5ex}\\
0 & 0 & 0
\end{array} \right)
V_V(x') +
\left( \begin{array}{ccc}
1 & 0 & 0 \\
0 & 1 & 0 \rule[-1.7ex]{0ex}{4.5ex}\\
0 & 0 & 1
\end{array} \right)
V_S(x') \right]
\left(\begin{array}{c} \psi_1(x') \\\rule[-1.7ex]{0ex}{4.5ex}
\psi_2(x') \\ \psi_3(x')
\end{array}\right).
\label{qq23}
\end{equation}
In this formula $E=M/2$ and $V_V(x)$ and $V_S(x)$ denote the vector
and scalar part of the q\=q potential. The Feynman gauge is assumed.
We first consider the case of a fixed coupling constant $\alpha$.
Assuming that $\psi_i(x) = c_i x^{\gamma_i}$ for $x \rightarrow 0$
one can reduce in leading order in $x$ the BSLT equation to
\begin{equation}
\left\{
\begin{array}{ll}
\displaystyle \rule[-2ex]{0ex}{6ex}
-mc_1 x^{\gamma_1} + E c_2x^{\gamma_2} +
c_3(\gamma_3+1)x^{\gamma_3-1} &
\displaystyle
=
-\frac{\alpha}{4}\left[
-2mc_1 x^{\gamma_1} I^{(0)}(\gamma_1)
+4Ec_2 x^{\gamma_2} I^{(0)}(\gamma_2) \right]
\\
\displaystyle \rule[-3ex]{0ex}{8ex}
Ec_1 x^{\gamma_1} - m c_2x^{\gamma_2} &
\displaystyle
=
-\frac{\alpha}{4}\left[
-2Ec_1 x^{\gamma_1} I^{(0)}(\gamma_1)
+4mc_2 x^{\gamma_2} I^{(0)}(\gamma_2) \right]
\\
\displaystyle \rule[-2ex]{0ex}{6ex}
c_1 (\gamma_1-1) x^{\gamma_1-1} - m c_3x^{\gamma_3} &
\displaystyle
=
-\frac{\alpha}{4}\left[
-2c_1 (\gamma_1-2)x^{\gamma_1-1} I^{(1)}(\gamma_1-1)
\right],
\end{array} \right.
\label{qq24}
\end{equation}
where the $I^{(\ell)}$ follow from
\begin{equation}
\frac{1}{m} \int_0^\infty \! dx'\ xx'\
Z^{(\ell)}(x,x')\ x'^{\gamma-1}
=x^\gamma I^{(\ell)}(\gamma),
\end{equation}
which is well-defined if $x\rightarrow 0$ and $|\gamma|<1$.
Assuming that
Eq.~(\ref{qq24}) does not solely determine the energy eigenvalue E,
we may exclude the case that $\gamma_1=\gamma_2$. As a result we
find that the exponents satisfy $\gamma_2<\gamma_1$,
$\gamma_3=\gamma_2+1$ and $\alpha^{-1} = I^{(0)}(\gamma_2)$. Taking
the explicit form of
\begin{equation}
Z^{(0)}(x,x') = \frac{m}{\pi xx'}
\left[ K_0(m|x-x'|) - K_0(m|x+x'|) \rule[-1.7ex]{0ex}{4.5ex}\right]
\end{equation}
we get
\begin{equation}
I^{(0)}(\gamma) =
\frac{1}{\pi} \int_0^1 \frac{d\eta}{\eta}
\ln \left(\frac{1+\eta}{1-\eta}\right)
\left( \eta^{\gamma} + \eta^{-\gamma} \right) =
\frac{1}{\gamma} \tan{\frac{\pi\gamma}{2}}.
\end{equation}
Explicit expressions for general $I^{(\ell)}$ are given in
Ref.~\cite{murota}. Hence
\begin{eqnarray}
\lefteqn{
\alpha = \frac{1}{I^{(0)}(\gamma_2)} \leq \frac{2}{\pi}.}
\hspace{17em}&&\mbox{(BSLT)}
\label{qq32}
\end{eqnarray}
This may be compared  with the asymptotic behavior of the
BSE~\cite{soper}, where
\begin{eqnarray}
\lefteqn{
\gamma_2 =\gamma_3 - 1 = \sqrt{1-\frac{4\alpha}{\pi}}
\ \ \rightarrow\ \
\alpha \leq \frac{\pi}{4},}
\hspace{17em}&&\mbox{(BSE)}
\rule[-3ex]{0ex}{3ex}
\\
\lefteqn{
\gamma_1 = \gamma_4 =  \sqrt{5-2\sqrt{4+\frac{\alpha}{\pi}
\left(1+\frac{\alpha}{\pi}\right)}}.}
\hspace{17em}&&\mbox{(BSE)}
\end{eqnarray}
Here $\gamma_4$ is the exponent of the $^3P_0^o$ component which
does not decouple in the BSE. Note that the dependence of $\gamma_2$
on $\alpha$ is to first order identical for the BSLT equation as for
the BSE.

The ET uv behavior can be found in a similar way and was already
discussed in Ref.~\cite{peter}. Here again $\gamma_2 < \gamma_1$,
$\gamma_3=\gamma_2+1$ and
\begin{eqnarray}
\lefteqn{
\alpha = \frac{1}{2}\gamma_2 (1-\gamma_2) I^{(0)}(1-\gamma_2)
\leq \frac{1}{\pi}.}
\hspace{17em}\mbox{(ET)}
\label{qq33}
\end{eqnarray}
If the Feynman gauge is replaced by the Coulomb gauge then according
to Eq.~(\ref{qq22}) we have to replace the diagonal matrix
$(-2,4,0)$ in Eq.~(\ref{qq23}) by the diagonal matrix $(-1,3,-1)$
and one gets for the BSLT and ET equations
\begin{eqnarray}
\lefteqn{\alpha =
\frac{4}{3I^{(0)}(\gamma_2)} \leq \frac{8}{3\pi},}
\hspace{17em}&&\mbox{(BSLT)}
\label{qq36} \\
\lefteqn{\alpha =
\frac{2}{3}\gamma_2 (1-\gamma_2) I^{(0)}(1-\gamma_2)
\leq \frac{4}{3\pi}.}
\hspace{17em}&&\mbox{(ET)}
\rule{0ex}{5ex}
\label{qq34}
\end{eqnarray}
So the short distance behavior is less singular in the Coulomb gauge
than in the Feynman gauge. This can be understood from the
suppression of the spin-spin interaction in the Coulomb gauge.  The
same singular behavior has been found by Murota~\cite{murota} for
the Salpeter equation, which is obtained from the ET equation by
removing the $|\!+-\rangle$ and $|\!-+\rangle$ components.

If the short distance attraction of the Coulomb-like potential is
weakened by taking a running coupling constant $\alpha(x) \sim
\ln^{-1}(x)$ then the singular short distance behavior of the
$\psi_i$ is also weakened to $\psi(x)=c_i
|\ln(x)|^{\gamma_i}$. However, to construct numerically the
solutions to the quasi-potentials equations it is not necessary to
know the expressions for the $\gamma_i$. The equations can be solved
by assuming the singular behavior associated with a fixed coupling
constant which is equal to the running coupling constant at a
typical small distance.

\figure{
\label{fig1}
The meson wave function at zero relative distance written as a loop
integral over the quark (quasi-) propagators and the potential, with
additional corrections. If this loop integral is divergent, then the
wave function is divergent at $x=0$.}

\figure{
\label{fig2}
A physical bound state $^1\!S_0$ solution (a) at $M=1.23\
\mbox{GeV}$ and an unphysical continuum $^1\!S_0$ solution (b) at
$M=1.30\ \mbox{GeV}$
obtained from the BSLT equation, with $m=0.2\ \mbox{GeV},
\ \kappa=0.2\ \mbox{GeV}^2,\ \alpha=c=\epsilon=0$.}

\figure{
\label{fig3}
The masses of the ground state and the first few excitations of the
$^1S_0$ systems are shown as a function of the fixed coupling
constant $\alpha$ (calculated with $m=5\ \mbox{GeV},\ \kappa=0.2\
\mbox{GeV}^2,
\ \varepsilon=c=0$).
The solid lines are for BSLT, broken lines for ET, and dotted for
Schr\"odinger. }

\figure{
\label{fig4}
The masses of the ground state and the first few excitations of the
$^1S_0$ system are shown as a function of the constant term $c$ in
the q\=q potential (calculated with $m=0.2\ \mbox{GeV},
\ \kappa=0.2\ \mbox{GeV}^2,\ \varepsilon=\alpha=0$).
States with negative mass represent anti-mesons. The solid lines are
for BSLT, broken lines for ET. }

\figure{
\label{fig5}
The masses of the ground state and the first few excitations of the
$^1S_0$ systems are shown as a function of the quark mass
(calculated with $c=\varepsilon=\alpha=0$), but with the ratio
$\kappa/m^2 = 0.2$ fixed. Under this condition the Schr\"odinger
levels are independent of $m$; they are indicated by the arrows on
the right of the figure. The solid lines are for BSLT, broken lines
for ET. }

\figure{
\label{fig6}
The masses of the ground state and the first few excitations of the
$^1S_0$ systems in the BSLT equation are shown as a function of the
mass $m_1$ of the first quark (calculated $m_2=0.2\ \mbox{GeV},\
\kappa=0.2\ \mbox{GeV}^2, \ c=\varepsilon=\alpha=0$).}

\figure{
\label{fig7}
Regge-slopes in the BSLT model. The squared meson masses as a
function of the total angular momentum $J$ of the unexcited mesons
(calculated with $m=0.25\ \mbox{GeV},\ \alpha=0,
\ c=-1.0\ \mbox{GeV}$ and $\kappa=0.33\ \mbox{GeV}^2$).
Solid lines show the masses for a pure scalar confinement
($\varepsilon = 0$), broken lines for mixture of vector and mainly
scalar confinement in the Feynman gauge ($\varepsilon = 0.15$), and
dotted lines for the same mixture using the Coulomb gauge. The lines
labeled $\pi$ represent the $\pi$ family [$P=(-)^{J\!+\!1}$ and
$C=(-)^J$], and similarly for the $\rho$ family
[$P=C=(-)^{J\!+\!1}$] and the $a_1$ family [$P=C=(-)^J$]. The
corresponding Regge-slopes are listed in Table~\ref{table1}.}

\figure{
\label{fig8}
Identical to Fig.~\ref{fig7}, but now for the ET equation.}

\mediumtext
\begin{table}
\caption{Slope $\beta$ (in $\mbox{GeV}^2$)
of the Regge-trajectories shown in Fig.~\ref{fig7} and
Fig.~\ref{fig8}. Listed values follow from a fit from $J=2$ to
$J=6$.}
\begin{tabular}{lllll}
&family & $\varepsilon=0$ & $\varepsilon=0.15$ & $\varepsilon=0.15$
\\
\makebox[15mm][l]{\ }&\makebox[20mm][l]{\ }&\makebox[20mm][l]{\ }&
\makebox[20mm][l]{Feynman}&\makebox[20mm][l]{Coulomb}\\
\tableline
BSLT
   &$\pi$  & 0.69    & 1.20       & 1.06   \\
   &$\rho$ & 0.69    & 1.13       & 1.01   \\
   &$a_1$  & 0.66    & 1.12       & 1.01   \\ ET\rule{0ex}{4ex}
   &$\pi$  & 0.67    & 0.47       & 0.64   \\
   &$\rho$ & 0.69    & 1.04       & 0.99   \\
   &$a_1$  & 0.65    & 0.92       & 0.83   \\
\end{tabular}
\label{table1}
\end{table}

\mediumtext
\begin{table}
\caption{Parameters for the BSLT and ET model (in units of GeV).}
\begin{tabular}{llll}
\rule{1.5cm}{0cm}& BSLT & BSLT& ET  \\
& Feynman gauge & Coulomb gauge & Coulomb gauge \\
\tableline
 $m_{u,d}$ & 0.200 & 0.250 & 0.250 \\
 $m_s$     & 0.404 & 0.447 & 0.390 \\
 $m_c$     & 1.715 & 1.779 & 1.719 \\
 $m_b$     & 5.121 & 5.199 & 5.096 \\

 $\kappa$      & 0.33  & 0.33  & 0.33 \rule{0ex}{4ex}\\
 $c$           &-0.8   &-1.0   &-1.0   \\
 $\varepsilon$&0.2 & 0.25  & 0.2 \\
 $\alpha_{sat}$&0.8& 0.8   & 0.8   \\ running type  &$\alpha_I,\
\mu=1$&$\alpha_I,\ \mu=1$&$\alpha_{II}$\\
\end{tabular}
\label{table2}
\end{table}

\mediumtext
\begin{table}
\caption{
The meson mass spectrum (in GeV). The experimental data are from the
particle data group~\cite{part}. Numbers between brackets need
confirmation. Only experimental errors larger than 0.01 GeV are
given. The column labeled $N^{2S+1}\!L_J$ displays the quantum
numbers of the main component of the wave function. Underlined
values have been fitted. The most satisfying results are found from
the BSLT using the Feynman gauge.}
\begin{tabular}{llllllll}
quark  & meson & $J^{PC}$ & $N^{2S+1}\!L_J$& Observed
&BSLT&BSLT&ET\\ content&       &  & &mass&Feynman&Coulomb&Coulomb\\
\tableline
u\=u, d\=d
 &$\pi$       &$0^{-+}$&$1\;^1S_0$& 0.135          & 0.439&
0.521&0.134 \\
 &$\pi'$      &$0^{-+}$&$2\;^1S_0$& 1.30$\pm$ 0.10 & 1.441&
1.424&1.361 \\
 &$\pi''$     &$0^{-+}$&$3\;^1S_0$&(1.77$\pm$ 0.03)& 2.246&
2.193&1.978 \\
\rule{0ex}{4ex}
 &$\rho$      &$1^{--}$&$1\;^3S_1$& 0.768          & 0.798&
0.796&0.838 \\
 &$\rho'$     &$1^{--}$&$1\;^3D_1$& 1.47$\pm$ 0.03 & 1.454&
1.470&1.613 \\
 &$\rho''$    &$1^{--}$&$2\;^3S_1$& 1.70$\pm$ 0.02 & 1.653&
1.594&1.649 \\
 &$\rho'''$   &$1^{--}$&$2\;^3D_1$&(2.14$\pm$ 0.03)& 2.185&
2.180&2.230 \\
 &$\rho''''$  &$1^{--}$&$3\;^3S_1$&                & 2.367&
2.308&2.248 \\
\rule{0ex}{4ex}
 &$b_1$       &$1^{+-}$&$1\;^1P_1$& 1.23           & 1.091&
1.136&1.196 \\
 &$a_0$       &$0^{++}$&$1\;^3P_0$& 0.983          & 0.993&
1.001&1.186 \\
 &$a_1$       &$1^{++}$&$1\;^3P_1$& 1.26$\pm$ 0.03 & 1.126&
1.142&1.249 \\
 &$a_2$       &$2^{++}$&$1\;^3P_2$& 1.318          & 1.297&
1.277&1.311 \\
\rule{0ex}{4ex}
 &$\pi_2$     &$2^{-+}$&$1\;^1D_2$& 1.67$\pm$ 0.02 & 1.524&
1.552&1.569 \\
 &$b_3$       &$3^{+-}$&$1\;^1F_3$&                & 1.886&
1.896&1.820 \\
 &$\pi_4$     &$4^{-+}$&$1\;^1G_4$&                & 2.205&
2.198&2.011 \\
 &$b_5$       &$5^{+-}$&$1\;^1H_5$&                & 2.494&
2.471&2.164 \\
 &$\pi_6$     &$6^{-+}$&$1\;^1I_6$&                & 2.761&
2.722&2.236 \\
\rule{0ex}{4ex}
 &$\rho_3$    &$3^{--}$&$1\;^3D_3$& 1.69           & 1.689&
1.654&1.660 \\
 &$a_4$       &$4^{++}$&$1\;^3F_4$&(2.04$\pm$ 0.03)& 2.021&
1.973&1.946 \\
 &$\rho_5$    &$5^{--}$&$1\;^3G_5$&(2.35)          & 2.314&
2.253&2.191 \\
 &$a_6$       &$6^{++}$&$1\;^3H_6$&(2.45$\pm$ 0.13)& 2.583&
2.507&2.407 \\
\rule{0ex}{4ex}
 &$\rho_2$    &$2^{--}$&$1\;^3D_2$&                & 1.547&
1.562&1.624 \\
 &$a_3$       &$3^{++}$&$1\;^3F_3$&(2.08$\pm$ 0.04)& 1.897&
1.903&1.904 \\
 &$\rho_4$    &$4^{--}$&$1\;^3G_4$&                & 2.202&
2.199&2.133 \\
 &$a_5$       &$5^{++}$&$1\;^3H_5$&                & 2.479&
2.465&2.330 \\
 &$\rho_6$    &$6^{--}$&$1\;^3I_6$&                & 2.732&
2.709&2.503 \\
\rule{0ex}{4ex}
d\=s
 &$K$         &$0^{-}$ &$1\;^1S_0$& 0.498          & 0.593&
0.660&0.350 \\
 &$K'$        &$0^{-}$ &$2\;^1S_0$&(1.46)          & 1.560&
1.530&1.457 \\
 &$K''$       &$0^{-}$ &$3\;^1S_0$&(1.83)          & 2.326&
2.268&2.069 \\
\rule{0ex}{4ex}
 &$K^*$       &$1^{-}$ &$1\;^3S_1$& 0.896          &
                           \underline{0.896}&\underline{0.896}&0.910
\\
 &$K^{*'}$    &$1^{-}$ &$1\;^3D_1$& 1.41           & 1.584&
1.599&1.692 \\
 &$K^{*''}$   &$1^{-}$ &$2\;^3S_1$& 1.71$\pm$0.02  & 1.727&
1.680&1.700 \\
\rule{0ex}{4ex}
 &$K_1$       &$1^{+}$ &$1\;^1P_1$& 1.27           & 1.232&
1.257&1.282 \\
 &$K_0^*$     &$0^{+}$ &$1\;^3P_0$& 1.43           & 1.113&
1.129&1.262 \\
 &$K_1'$      &$1^{+}$ &$1\;^3P_1$& 1.40           & 1.261&
1.268&1.333 \\
 &$K_2^*$     &$2^{+}$ &$1\;^3P_2$& 1.425          & 1.389&
1.373&1.376 \\
\rule{0ex}{4ex}
 &$K_2$       &$2^{-}$ &$1\;^1D_2$& 1.77           & 1.655&
1.669&1.672 \\
 &$K_3$       &$3^{+}$ &$1\;^1F_3$&(2.32$\pm$0.02) & 2.005&
2.008&1.943 \\
 &$K_4$       &$4^{-}$ &$1\;^1G_4$&(2.49$\pm$0.02) & 2.331&
2.304&2.153 \\
 &$K_5$       &$5^{+}$ &$1\;^1H_3$&                & 2.590&
2.571&2.326 \\
\rule{0ex}{4ex}
 &$K_3^*$     &$3^{-}$ &$1\;^3D_3$& 1.77           & 1.776&
1.749&1.726 \\
 &$K_4^*$     &$4^{+}$ &$1\;^3F_4$& 2.05           & 2.107&
2.069&2.018 \\
 &$K_5^*$     &$5^{-}$ &$1\;^3G_5$&(2.38$\pm$0.03) & 2.399&
2.350&2.271 \\
\rule{0ex}{4ex}
u\=c
 &$D$         &$0^{-}$ &$1\;^1S_0$& 1.864          & 1.868&
1.912&1.855 \\
 &$D^*$       &$1^{-}$ &$1\;^3S_1$& 2.007          & 2.015&
2.032&2.058 \\
\rule{0ex}{4ex}
 &$D_1$       &$1^{+}$ &$1\;^1P_1$& 2.42           & 2.388&
2.404&2.423 \\
 &$D_0^*$     &$0^{+}$ &$1\;^3P_0$&                & 2.321&
2.327&2.632 \\
 &$D_1'$      &$1^{+}$ &$1\;^3P_1$&                & 2.415&
2.420&2.494 \\
 &$D_2^*$     &$2^{+}$ &$1\;^3P_2$& 2.459          & 2.458&
2.461&2.474 \\
\rule{0ex}{4ex}
s\=c
 &$D_s$       &$0^{-}$ &$1\;^1S_0$& 1.969          & 1.952&
1.983&1.939 \\
 &$D_s^*$     &$1^{-}$ &$1\;^3S_1$&(2.110)         & 2.104&
2.110&2.115 \\
\rule{0ex}{4ex}
 &$D_{s1}$    &$1^{+}$ &$1\;^1P_1$& 2.537          & 2.500&
2.509&2.491 \\
 &$D_{s0}^*$  &$0^{+}$ &$1\;^3P_1$&                & 2.427&
2.419&2.510 \\
 &$D_{s1}'$   &$1^{+}$ &$1\;^3P_1$&                & 2.516&
2.502&2.550 \\
 &$D_{s2}^*$  &$2^{+}$ &$1\;^3P_2$&(2.564)         & 2.569&
2.559&2.547 \\
\rule{0ex}{4ex}
c\=c
 &$\eta_c$    &$0^{-+}$&$1\;^1S_0$& 2.980          & 2.969&
2.988&2.883 \\
 &$\eta_c'$   &$0^{-+}$&$2\;^1S_0$&(3.59)          & 3.742&
3.713&3.680 \\
\rule{0ex}{4ex}
 &$J/\psi$    &$1^{--}$&$1\;^3S_1$& 3.097          &
               \underline{3.096}&\underline{3.098}&\underline{3.097}
\\
 &$\psi'$     &$1^{--}$&$2\;^3S_1$& 3.686          & 3.810&
3.779&3.789 \\
 &$\psi''$    &$1^{--}$&$1\;^3D_1$& 3.770          & 3.873&
3.854&3.872 \\
 &$\psi'''$   &$1^{--}$&$3\;^3S_1$& 4.04           & 4.370&
4.325&4.308 \\
 &$\psi^{iv}$ &$1^{--}$&$2\;^3D_1$& 4.16$\pm$ 0.02 & 4.409&
4.372&4.363 \\
 &$\psi^v$    &$1^{--}$&$3\;^3S_1$& 4.42           & 4.860&
4.805&4.744 \\
\rule{0ex}{4ex}
 &$h_{c1}$    &$1^{+-}$&$1\;^1P_1$& 3.526$^{\rm a}$& 3.517&
3.513&3.486 \\
 &$\chi_{c0}$ &$0^{++}$&$1\;^3P_0$& 3.415          & 3.461&
3.442&3.486 \\
 &$\chi_{c1}$ &$1^{++}$&$1\;^3P_1$& 3.511          & 3.526&
3.510&3.521 \\
 &$\chi_{c2}$ &$2^{++}$&$1\;^3P_2$& 3.556          & 3.572&
3.559&3.553 \\
\rule{0ex}{4ex}
d\=b
 &$B$         &$0^{-}$ &$1\;^1S_0$& 5.279          & 5.302&
5.331&5.349 \\
 &$B^*$       &$1^{-}$ &$1\;^3S_1$& 5.324          & 5.360&
5.383&5.391 \\
\rule{0ex}{4ex}
 &$B_1$       &$1^{+}$ &$1\;^1P_1$&                & 5.741&
5.756&5.741 \\
 &$B_0^*$     &$0^{+}$ &$1\;^3P_0$&                & 5.714&
5.717&5.800 \\
 &$B_1'$      &$1^{+}$ &$1\;^3P_1$&                & 5.760&
5.766&5.817 \\
 &$B_2^*$     &$2^{+}$ &$1\;^3P_2$&                & 5.770&
5.781&5.760 \\
\rule{0ex}{4ex}
s\=b
 &$B_s$       &$0^{-}$ &$1\;^1S_0$&(5.38$\pm$0.03) & 5.371&
5.383&5.390 \\
 &$B_s^*$     &$1^{-}$ &$1\;^3S_1$&(5.43$\pm$0.03) & 5.434&
5.443&5.441 \\
\rule{0ex}{4ex}
 &$B_{s1}$    &$1^{+}$ &$1\;^1P_1$&                & 5.839&
5.841&5.805 \\
 &$B_{s0}^*$  &$0^{+}$ &$1\;^3P_0$&                & 5.802&
5.789&5.848 \\
 &$B_{s1}'$   &$1^{+}$ &$1\;^3P_1$&                & 5.846&
5.838&5.865 \\
 &$B_{s2}^*$  &$2^{+}$ &$1\;^3P_2$&                & 5.869&
5.866&5.827 \\
\rule{0ex}{4ex}
c\=b
 &$B_c$       &$0^{-}$ &$1\;^1S_0$&                & 6.260&
6.260&6.228 \\
 &$B_c^*$     &$1^{-}$ &$1\;^3S_1$&                & 6.331&
6.329&6.336 \\
\rule{0ex}{4ex}
 &$B_{c1}$    &$1^{+}$ &$1\;^1P_1$&                & 6.760&
6.754&6.719 \\
 &$B_{c0}^*$  &$0^{+}$ &$1\;^3P_1$&                & 6.724&
6.702&6.724 \\
 &$B_{c1}'$   &$1^{+}$ &$1\;^3P_1$&                & 6.767&
6.745&6.742 \\
 &$B_{c2}^*$  &$2^{+}$ &$1\;^3P_2$&                & 6.794&
6.781&6.758 \\
\rule{0ex}{4ex}
b\=b
 &$\eta_b$    &$0^{-+}$&$1\;^1S_0$&                & 9.401& 9.402&
9.350 \\
 &$\eta_b'$   &$0^{-+}$&$2\;^1S_0$&                &10.067&10.047&
9.980 \\
\rule{0ex}{4ex}
 &$\Upsilon$    &$1^{--}$&$1\;^3S_1$& 9.460        &
               \underline{9.460}&\underline{9.459}&\underline{9.460}
\\
 &$\Upsilon'$   &$1^{--}$&$2\;^3S_1$& 10.023
&10.099&10.081&10.039 \\
 &$\Upsilon''$  &$1^{--}$&$1\;^3D_1$& 10.355
&10.206&10.187&10.138 \\
 &$\Upsilon'''$ &$1^{--}$&$3\;^3S_1$& 10.58
&10.556&10.532&10.467 \\
&$\Upsilon^{iv}$&$1^{--}$&$2\;^3D_1$& 10.87
&10.629&10.603&10.533 \\
 &$\Upsilon^v$  &$1^{--}$&$4\;^3S_1$& 11.02
&10.943&10.911&10.827 \\
\rule{0ex}{4ex}
 &$h_{b1}$    &$1^{+-}$&$1\;^1P_1$&                & 9.881& 9.879&
9.823 \\
 &$\chi_{b0}$ &$0^{++}$&$1\;^3P_0$& 9.860          & 9.862& 9.843&
9.831 \\
 &$\chi_{b1}$ &$1^{++}$&$1\;^3P_1$& 9.892          & 9.890& 9.876&
9.845 \\
 &$\chi_{b2}$ &$2^{++}$&$1\;^3P_2$& 9.913          & 9.911& 9.901&
9.863 \\
\rule{0ex}{4ex}
 &$h_{b1}'$   &$1^{+-}$&$2\;^1P_1$&
&10.383&10.363&10.288 \\
 &$\chi_{b0}'$&$0^{++}$&$2\;^3P_0$& 10.232
&10.363&10.335&10.289 \\
 &$\chi_{b1}'$&$1^{++}$&$2\;^3P_1$& 10.255
&10.384&10.360&10.301 \\
 &$\chi_{b2}'$&$2^{++}$&$2\;^3P_2$& 10.268
&10.400&10.379&10.316 \\
\end{tabular}
\label{table3}
\tablenotes{$^{\rm a}$ Ref.~\cite{e760}.}
\end{table}

\narrowtext
\begin{table}
\caption{
16-component vector qq-basis states and corresponding $4\times 4$
matrix q\={q}-basis states. The angular dependence of the wave
functions of the q\=q-triplet states must determine to which
qq-triplet state they correspond. For a correct normalization all
matrix states should be multiplied by $1/2$; furthermore, states
proportional to $P$ or $p$ need an extra factor $1/M$ and $1/|p|$,
respectively. The correspondence is only valid in the cm system and
$p_0=0$ is assumed.}
\begin{tabular}{ll} qq-state & q\={q}-state  \\
\tableline
 $ ^1J_J^s $&$ -\gamma_5 \not\! P$\\
 $ ^1J_J^a $&$ \gamma_5          $\\
 $ ^1J_J^e $&$ -1                $\\
 $ ^1J_J^o $&$ \not\! P          $\\
 $ ^3J_J^s,\ ^3(J\!-\!1)_J^s,\ ^3(J\!+\!1)_J^s $&
 $ -\not\! p $\\
 $ ^3J_J^a,\ ^3(J\!-\!1)_J^a,\ ^3(J\!+\!1)_J^a $&
 $ i\sigma_{\mu\nu} P^\mu p^\nu $\\
 $ ^3J_J^e,\ ^3(J\!-\!1)_J^e,\ ^3(J\!+\!1)_J^e $&
 $ -\gamma_5 i\sigma_{\mu\nu} P^\mu p^\nu $\\
 $ ^3J_J^o,\ ^3(J\!-\!1)_J^o,\ ^3(J\!+\!1)_J^o $&
 $ \gamma_5 \not\! p $ \\
\end{tabular}
\label{table4}
\end{table}

\end{document}